\documentclass[aps,twocolumn,superscriptaddress]{revtex4}

\usepackage{bm,amsmath,color,multirow,graphicx}

\begin{document}
\hspace{1pt}

\hspace{1pt}

\rightline{\bf{YITP-15-66}}

\hspace{1pt}

\title{Surface electronic state of superconducting topological crystalline insulator}

\author{Tatsuki Hashimoto}
\affiliation{Department of Applied Physics, Nagoya University, Nagoya 464-8603, Japan}
\author{Keiji Yada}
\affiliation{Department of Applied Physics, Nagoya University, Nagoya 464-8603, Japan}
\author{Masatoshi Sato}
\affiliation{Yukawa Institute for Theoretical Physics, Kyoto University, Kyoto 606-8502, Japan}
\author{Yukio Tanaka}
\affiliation{Department of Applied Physics, Nagoya University, Nagoya 464-8603, Japan}

\date{\today}

\begin{abstract}
We study the surface state of a doped topological crystalline insulator in the superconducting state. 
Motivated by Sn$_{1-x}$In$_{x}$Te, we consider fully-gapped pair potentials and calculate the surface spectral function. 
It is found that mirror-protected zero-energy surface Andreev bound states (SABSs) appear at the (001) surface. The gapless points of these SABSs appear on the mirror symmetric line on the surface Brillouin zone while the positions of the gapless points depend on the chemical potential. In addition, due to the presence of the Dirac surface states in the normal state, the dispersion of the SABSs drastically changes with the chemical potential.
\end{abstract}

\pacs{pacs}
\maketitle

\section{Introduction}
Surface Andreev bound state (SABS) is an important physics in superconductors.\cite{ABS1,ABS2,Deutscher}. SABSs appear on the surface of an unconventional superconductor with a pair potential that changes its sign on the Fermi surface. \cite{ABS,ABSb,Hu,TK95,ABS1} These SABSs have been detected  by tunneling spectroscopy measurements in cuprate \cite{TK95,ABS1} and Sr$_2$RuO$_4$ \cite{Yamashiro,Kashiwaya11}.
The topological origin of 
SABSs has been clarified based on the topological invariants defined in the bulk Hamiltonian. 
Because of bulk-boundary correspondence, superconductors with gapless SABSs are now called topological superconductors \cite{tanaka12,alicea12}, and the concept of topological superconductivity has been expanding widely \cite{Law,Brydon1,Schnyder2011,SchnyderBrydon,STYY11,TMYYS10,YSTY10,STF09,lutchyn10,oreg10,alicea10}.

On the other hand, topological insulator has been of great interest because of its novel surface state protected by time-reversal symmetry \cite{RevModPhys.82.3045,RevModPhys.83.1057,doi:10.7566/JPSJ.82.102001}. The superconducting topological insulator (STI), which is a superconductor realized in a doped topological insulator, has been also attracting a great research interest following the first observation of zero-bias conductance peak that indicates the existence of SABSs in Cu$_x$Bi$_2$Se$_3$ \cite{Sasaki}. And now, determination of the pairing symmetry of Cu$_x$Bi$_2$Se$_3$ is recognized as an important problem \cite{Hor,Sasaki2015206,0953-2048-27-10-104002,doi:10.7566/JPSJ.82.044704,doi:10.7566/JPSJ.83.064705,Nagai1,Nagai2,Nagai3,Nagai4,Yip,FuBerg,PhysRevB.90.100509,Kriener1,Kriener2,Bay,Zocher,PhysRevB.90.184512}. In the STI, theoretical studies have revealed that the cone shape SABS becomes twisted dispersion due to the mixing with the surface state in the normal state \cite{PhysRevB.85.180509,PhysRevLett.108.107005,PhysRevB.83.134516}. The structural change of the dispersion is considered as an important property for the interpretation of the tunneling spectroscopy. The existence of peculiar SABSs has been also predicted in the doped topological materials such as Dirac semimetal and Weyl semimetal \cite{2015arXiv150407408K,PhysRevLett.114.096804,PhysRevB.92.035153}.

Recently, a new type of topological material, topological crystalline insulator (TCI), has been proposed \cite{PhysRevLett.106.106802}. TCIs are the materials that have surface states protected by point group symmetry, e.g. mirror reflection symmetry and rotational symmetries. Up to now, SnTe, Pb$_{1-x}$Sn$_x$Te and Pb$_{1-x}$Sn$_x$Se heve been revealed to be the TCI with non zero mirror Chern number \cite{NatCommFu,SnTe_ARPES,PbSnSe_ARPES,PbSnTe_ARPES}. They host double Dirac cones on the (001) surface protected by mirror reflection symmetry [see Fig.\ref{fig_cry}(c)]. Since the surface states of the TCIs are protected by point group symmetry, the Dirac cone can be controlled by the electronic field or strain as well as the magnetic field. The additional tunability of the surface state can be applicable to the high-speed topological transistor \cite{TCI_app,PhysRevLett.112.046801,TCI_app_strain}.
For TCIs, superconductivity has been reported in In-doped SnTe \cite{InSnTe_SC}. Point contact measurements in superconducting Sn$_{1-x}$In$_{x}$Te show zero-bias conductance peak that may originate from SABSs \cite{PhysRevLett.109.217004}. In other words, this material is a candidate of a topological superconductor. Furthermore, it is expected that superconducting TCIs (STCIs) exhibit a new type of SABS since the TCIs host the unique surface Dirac cone in the normal state.

Using the similarity of low energy electronic state to Cu$_x$Bi$_2$Se$_3$, Sasaki $et.$ $al.$ have investigated the superconducting state of Sn$_{1-x}$In$_{x}$Te with an effective model for around each $L$ point (see Fig. \ref{fig_cry}) \cite{PhysRevLett.109.217004}. However, different from Cu$_x$Bi$_2$Se$_3$ which has a single Fermi surface, Sn$_{1-x}$In$_{x}$Te has four hole pockets at around each $L$ point. Then, two hole pockets are projected to the same position in the surface Brillouin zone when we see the (001) surface. Therefore, it is inevitable to consider the overlapping of the Fermi surfaces to reveal the surface state properly.

In this paper, we study the surface states of the STCI using a tight-binding model which can describe the electronic states for the entire Brillouin zone. We introduce possible fully gapped pair potentials for the TCI and calculate the surface spectral function based on the recursive Green's function method. 
It is found that the STCI can host mirror-protected SABSs. These SABSs exhibit twisted dispersion due to the surface Dirac cone in the normal state. In contrast to STIs, the position of gapless points of SABS is not fixed to the time reversal invariant momenta but on the mirror invariant lines in the surface Brillouin zone. Thus, the position of the gapless points depends on the material parameters such as chemical potential. As a result, the dispersion of the SABSs in the STCI is not a zero-energy flat one which has been seen in STIs, but a more complicated one. To understand the topological nature of this material in more detail, we also calculate the mirror Chern number and construct the phase diagram in terms of the magnitudes of the pairing potential and the spin-orbit interaction. We find that the mirror Chern numbers $n_M$ takes three kinds of values, $n_M= 0$, $-2$ and $-4$, in STCI. Those values are different from those in STIs with $n_M= 0, -1, -2$.\\

This paper is organized as follows. In Sec. II, we review a tight-binding model proposed by Lent $et$. $al$. for IV-VI semiconductors and propose possible pair potentials for the tight-binding model. The topological nature of the TCI with possible pair potentials is discussed in Sec. III. In Sec. IV, we show our calculated results of spectral function for the (001) surface and discuss the evolution of the mirror-protected SABS. In Sec. V, we discuss the experimental situation and possibility of time-reversal breaking pairing. Finally, we summarize our results in Sec. VI.
\section{Model}
\begin{figure}[t]
\begin{center}
\includegraphics[width=8cm]{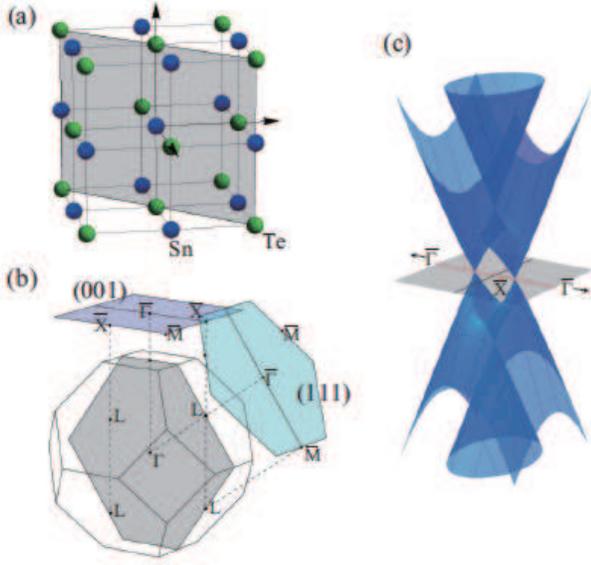}
\caption{(a)Crystal strcture of SnTe. (b) First Brillouin zone of a rock salt crystal and, (001) and (111) surface Brillouin zone. The shaded planes are (110) mirror invariant planes. (c) Schematic picture of the surface double Dirac cone on the (001) surface.}
\label{fig_cry}
\end{center}
\end{figure}
Before the introduction of a model, we briefly review the topological crystalline insulator SnTe \cite{NatCommFu}. In Fig.\ref{fig_cry}, we show the crystal structure (a) and, the first Brillouin zone and the surface Brillouin zone (b) of SnTe. The crystal of SnTe is the rock-salt structure, which belongs to the O$\rm _h$ point group. 
The band inversion occurs at four $L$ points in SnTe. 
For the TCIs, the (001) surface is particularly unique because two $L$ points are projected to the same $\bar{X}$ points on the surface Brillouin zone. 
In this case, two surface Dirac cones show up at the same $\bar{X}$ points with different energies. The Dirac cones mix with each other and thus have gap in general. However, on $\bar{\Gamma} -\bar{X}$ mirror-symmetric line, two Dirac cones remain gapless because each Dirac cone belongs to different mirror subspace. A schematic picture of the Dirac cone is shown in Fig. \ref{fig_cry} (c). These surface states have been observed by angle-resolved photoemission spectroscopy (ARPES) experiments \cite{SnTe_ARPES,PbSnTe_ARPES}. It is noted that even if In is doped, SnTe still has the gapless Dirac cones \cite{PhysRevLett.110.206804}. 

To describe the normal state of the TCI, we use a tight-binding model proposed by Lent $et$. $al.$ \cite{model}. This model Hamiltonian is a 36 $\times$ 36 matrix with nine orbitals, i.e. $s$, $p^3$, $d^5$, two spin degrees of freedom, and two sublattices (anion and cation). The Hamiltonian has the following nearest-neighbor hopping and the on-site spin-orbit interaction, which is given by
\begin{align}
\hat{H}_{{\rm SnTe}}=
\begin{pmatrix} 
\hat{H}_{s,s}&\hat{H}_{pc,s}^\dagger&\hat{H}_{pa,s}^\dagger&0&0\\
\hat{H}_{pc,s}&\hat{H}_{pc,pc}&\hat{H}_{pa,pc}^\dagger&0&\hat{H}_{da,pc}^\dagger\\
\hat{H}_{pa,s}&\hat{H}_{pa,pc}&\hat{H}_{pa,pa}&\hat{H}_{dc,pa}^\dagger&0\\
0&0&\hat{H}_{dc,pa}&\hat{H}_{dc,dc}&\hat{H}_{da,dc}^\dagger\\
0&\hat{H}_{da,pc}&0&\hat{H}_{da,dc}&\hat{H}_{da,da}
\end{pmatrix}.\label{eq_tbmodel}
\end{align}
Here, indices $a$ and $c$ denote anion and cation, respectively, and we take the basis as follows,
($c_{s,c,\uparrow}$,$c_{s,c,\downarrow}$,$c_{s,a,\uparrow}$,$c_{s,a,\downarrow}$,$c_{p_x,c,\uparrow}$,$c_{p_y,c,\uparrow}$,$c_{p_z,c,\uparrow}$,$c_{p_x,c,\downarrow}$,$c_{p_y,c,\downarrow}$,
$c_{p_z,c,\downarrow}$,$c_{p_x,a,\uparrow}$,$c_{p_y,a,\uparrow}$,$c_{p_z,a,\uparrow}$,$c_{p_x,a,\downarrow}$,$c_{p_y,a,\downarrow}$,$c_{p_z,a,\downarrow}$,$c_{d_{x^2-y^2},c,\uparrow}$,
$c_{d_{3z_2-r^2},c,\uparrow}$,$c_{d_{xy},c,\uparrow}$,$c_{d_{yz},c,\uparrow}$,$c_{d_{zx},c,\uparrow}$,$c_{d_{x^2-y^2},c,\downarrow}$,$c_{d_{3z^2-r^2},c,\downarrow}$,
$c_{d_{xy},c,\downarrow}$,$c_{d_{yz},c,\downarrow}$,$c_{d_{zx},c,\downarrow}$,$c_{d_{x^2-y^2},a,\uparrow}$,$c_{d_{3z^2-r^2},a,\uparrow}$,$c_{d_{xy},a,\uparrow}$,$c_{d_{yz},a,\uparrow}$,
$c_{d_{zx},a,\uparrow}$,$c_{d_{x^2-y^2},a,\downarrow}$,$c_{d_{3z^2-r^2},a,\downarrow}$,$c_{d_{xy},a,\downarrow}$,$c_{d_{yz},a,\downarrow}$,$c_{d_{zx},a,\downarrow}$).
See the original paper for the explicit form of the Hamiltonian and the parameters for SnTe \cite{model}.
The BdG Hamiltonian for the superconducting state is given by
\begin{align}
\hat{H}_{\rm BdG}=[\hat{H}_{\rm SnTe}-\mu\hat{I}]\hat{\tau}_z+\hat{\Delta}_i\hat{\tau}_x,
\label{BdG}
\end{align}
where $\hat{\tau}_i$ represents the Pauli matrix in the Nambu space, $\mu$ is the chemical potential, $\hat{I}$ is an identity matrix and $\hat{\Delta}_i$ is the pair potential. The basis is taken as $(c_{\alpha,\beta,\uparrow}, c_{\alpha,\beta,\downarrow}, -c^\dagger_{\alpha,\beta,\downarrow},c^\dagger_{\alpha,\beta,\uparrow})$, with $\alpha = s, p_x, p_y, p_z, d_{x^2-y^2}, d_{3z^2-r^2}, d_{xy}, d_{yz}, d_{zx}$ and $\beta = a, c$.\\

Up to now, almost all experiments for superconducting state in Sn$_{1-x}$In$_{x}$Te, e.g., specific heat \cite{PhysRevB.88.140502,PhysRevB.79.024520,PhysRevB.87.140507}, $\mu$-SR measurements \cite{PhysRevB.90.064508}, and thermal conductivity\cite{PhysRevB.88.014523}, have suggested the fully gapped superconductivity. 
One of the candidates of the fully-gapped superconducting state is $s$-wave pair potential, which is a spin-singlet even-parity on-site pair potential. This pair potential is essentially the same as BCS state, and therefore, the energy-gap structure is fully gapped.
Hereafter, we denote this pair potential as $\Delta_1$. 
Other than the above BCS state, one of the $p$-wave states realizes the fully gapped state.
For $p$-wave pair potential in the cubic lattice, there are four types of time-reversal-invariant pairing, $A_{1u}$, $E_{u}$, $T_{1u}$, and $T_{2u}$, as summarized in Refs. [\onlinecite{RevModPhys.63.239,PhysRevB.31.7114,PhysRevB.32.2935,PhysRevB.30.4000}]. The energy gap of $A_{1u}$ pairing is fully gapped. While the other pairings give point or line nodes. Thus, according to the experimental situation, we focus on $A_{1u}$ pair potential, and hereafter, we denote the $A_{1u}$ pair potential as $\Delta_2$. 
$\Delta_2$ is a spin-triplet odd-parity one, which is similar to that for the BW phase of $^3$He. In the $\Delta_2$ state, the Cooper pair is formed between the 
nearest neighbor sites, $i.e.$,  Sn and Te sites. Following, we assume that Cooper pairs are formed within the $p$ orbitals and ignore the $s$ and $d$ orbitals since low energy band is mainly composed of $p$ orbitals. For simplicity, we also assume that electrons in $p_i$ orbital ($i=x,y,z)$ form Cooper pair along the $i$-direction. These simplifications do not give a qualitative change of the results.
The explicit form of the pair potentials is given by
\begin{align}
\hat{\Delta}_1
&=\Delta
\sum_{i=x,y,z}
\hat{p}_i
\hat{s}_0\hat{\sigma}_0,
\label{d1}
\\
\hat{\Delta}_{2}
&=\Delta
\sum_{i=x,y,z}{
\sin \frac{k_i}{2} \hat{p}_i \hat{s}_i
}
\hat{\sigma}_x,
\label{d2}
\end{align}
where $\hat{s}_i$ and $\hat{\sigma}_i$ are the Pauli matrices of the spin and sublatice, respectively. In this basis, $\hat{s}_0$ ($\hat{s}_\mu$ [$\mu=x,y,z$)] denotes the spin singlet (spin triplet) and $\sigma_0$ ($\sigma_x$) denotes the intrasitre (intersite) pairing. $\hat{p}_i$ is the following matrix:
\begin{align}
\hat{p}_i=
\begin{pmatrix}
0&0&0&0&0&0&0&0&0\\
0&\delta_{ix}&0&0&0&0&0&0&0\\
0&0&\delta_{iy}&0&0&0&0&0&0\\
0&0&0&\delta_{iz}&0&0&0&0&0\\
0&0&0&0&0&0&0&0&0\\
0&0&0&0&0&0&0&0&0\\
0&0&0&0&0&0&0&0&0\\
0&0&0&0&0&0&0&0&0\\
0&0&0&0&0&0&0&0&0
\end{pmatrix},
\end{align}
in the basis, ($c_{s,\beta,\gamma}$,$c_{p_x,\beta,\gamma}$,$c_{p_y,\beta,\gamma}$,$c_{p_z,\beta,\gamma}$,$c_{d_{x^2-y^2},\beta,\gamma}$,
$c_{d_{3z^2-r^2},\beta,\gamma}$,$c_{d_{xy},\beta,\gamma}$,$c_{d_{yz},\beta,\gamma}$,$c_{d_{zx},\beta,\gamma}$),
 where $\beta=\rm a, c$ and $\gamma=\uparrow, \downarrow$.
The pair potentials also have the mirror reflection symmetry with respect to the (110) plane. $\Delta_1$ ($\Delta_2$) is even (odd) under the mirror reflection 
\begin{align}
\hat{{\cal M}}_{\rm (110)}\hat{\Delta}_1\hat{{\cal M}}_{\rm (110)}^\dagger&=\hat{\Delta}_1,\\
\hat{{\cal M}}_{\rm (110)}\hat{\Delta}_2\hat{{\cal M}}_{\rm (110)}^\dagger&=-\hat{\Delta}_2.
\end{align}
Here, $\hat{{\cal M}}_{(110)}$ is the (110) mirror operator given by
\begin{align}
\hat{{\cal M}}_{\rm (110)}=
\frac{i}{\sqrt{2}}
(\hat{s}_x-\hat{s}_y)
\otimes
\hat{\sigma_0}
\otimes
\hat{{\cal M}}_o,
\end{align}
where $\hat{{\cal M}}_o$ is a mirror operator for the orbital space.
The explcit form of $\hat{{\cal M}}_o$ is given by,
\begin{align}
\hat{{\cal M}}_o=
\begin{pmatrix}
1&0&0&0&0&0&0&0&0\\
0&0&1&0&0&0&0&0&0\\
0&1&0&0&0&0&0&0&0\\
0&0&0&1&0&0&0&0&0\\
0&0&0&0&-1&0&0&0&0\\
0&0&0&0&0&1&0&0&0\\
0&0&0&0&0&0&1&0&0\\
0&0&0&0&0&0&0&0&1\\
0&0&0&0&0&0&0&1&0
\end{pmatrix}.
\end{align}

Since $\Delta_1$ has an even-parity under the (110) mirror operation, 
$H_{\rm BdG}$ commutes with
\begin{align}
\hat{{\cal M}}_{\rm (110)}^{+}
=
\begin{pmatrix}
\hat{{\cal M}}_{\rm (110)}&0\\
0&\hat{{\cal M}}_{\rm (110)}^*
\label{mirror_BdG_p}
\end{pmatrix},
\end{align}
in the mirror-invariant plane.
On the other hand, $\Delta_2$ has an odd parity under the (110) mirror operation. Therefore, $H_{\rm BdG}$ can be block diagonalized under the diagonal basis of a mirror operator
\begin{align}
\hat{{\cal M}}_{\rm (110)}^{-}=
\begin{pmatrix}
\hat{{\cal M}}_{\rm (110)}&0\\
0&-\hat{{\cal M}}_{\rm (110)}^*
\label{mirror_BdG_m}
\end{pmatrix},
\end{align}
 We summarize the characters of the pair potentials in Table \ref{tab_pair}.
\\

\begin{table}
\begin{tabular}{ccccccc}
\hline\hline
Pair potential& O$\rm _h$ & spin &$I$&$M_{(110)}$&energy gap \\
\hline
$\Delta_{1}$&$A_{1g}$&singlet&$+$&$+$&full gap \\

$\Delta_{2}$&$A_{1u}$&triplet &$-$&$-$&full gap \\

\hline\hline
\end{tabular}
\caption{Possible pair potentials for the STCI. $\Delta_1$ ($\Delta_2$) is even (odd) under inversion operation. $\Delta_1$ ($\Delta_2$) is even (odd) under the (110) mirror reflection operation.}
\label{tab_pair}
\end{table}
\section{Topological nature}\label{sec_topo}
\begin{figure}[tb]
\begin{center}
\includegraphics[width=6cm]{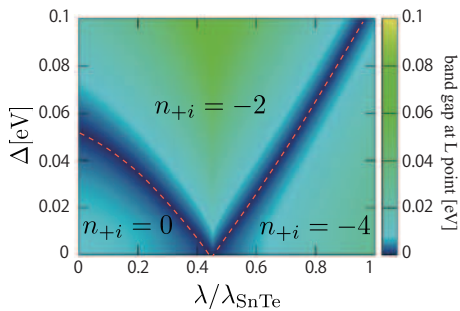}
\caption{Phase diagram of superconducting topological crystalline insulator with the odd-parity full-gapped pair potential $\Delta_2$ as a function of the magnitude of the pair potential $\Delta$ and spin-orbit interaction $\lambda/\lambda_{\rm SnTe}$. Color shows the magnitude of the energy gap at the $L$ point. Red dotted lines show gap closed line.}
\label{fig_phase}
\end{center}
\end{figure}
In this section, we clarify the topological nature of the STCI, based on the recently expanded topological periodic table that includes crystalline symmetries in addition to time-reversal and particle-hole symmetries \cite{PhysRevB.88.075142,PhysRevB.90.165114,PhysRevB.88.125129}. First of all, we briefly review the topological nature of the normal state. In terms of Altland-Zirnbauer (AZ) classification \cite{PhysRevB.55.1142,Kitaev2,Schnyder,PhysRevB.78.195125}, the TCI belongs to class AII, since it has the time-reversal symmetry for fermions. However, the corresponding $Z_2$ topological number is trivial, because band inversion occurs at an even number (four) of time-reversal invariant momenta. Thus, the time-reversal symmetry is not sufficient to explain the surface states in the TCI. Instead, the mirror reflection symmetry, which is specific to the crystal structure of the TCI, is responsible for the stability of the surface states in the TCI. Taking into account the mirror reflection symmetry, SnTe is classified as class AII with $U^{-}_+$ with $d=3$ and $d_{\parallel}=1$ in Ref. \cite{PhysRevB.90.165114}, because the mirror operator $\hat{{\cal M}}_{(110)}$ commutes with time-reversal operator $\hat{{\cal T}}$ and obeys $\hat{{\cal M}}_{(110)}^2=-1$. Therefore, its topological nature is characterized by the mirror Chern number
\begin{align}
n_{M}=\frac{n_{+i}-n_{-i}}{2}.
\end{align}
Here, $n_{\pm i}$ is the Chern number for mirror subsectors labeled with mirror eigenvalues $\pm i$. It has been known that the mirror Chern number for the TCI is $-2$ \cite{NatCommFu}.

Next, we see the topological nature in the superconducting state.
In terms of the AZ classification, the STCI with $\Delta_2$ belongs to class DIII. The superconducting state also possesses the mirror reflection symmetry. The mirror operator $\hat{{\cal M}}_{(110)}^-$ obeys $({\hat{{\cal M}}_{(110)}^-})^2=-1$ and commutes (anticomutes) with time reversal operator $\hat{{\cal T}}$ (chiral operator $\hat{{\cal C}}$). Thus, this system belongs to class DIII with $U^{-}_{+-}$ with $d=3$ and $d_{\parallel}=1$. In this case topological nature is characterized with two topological numbers, $Z \oplus Z$. One of the integer numbers is three-dimensional winding number. 
Although the exact calculation of the winding number is difficult because of the huge Hamiltonian matrix, its parity can be easily obtained by using Fermi surface criteria \cite{PhysRevB.81.220504}. In the present case, the Fermi surfaces enclose four time reversal invariant momenta, i.e., $L$ points. Thus, the winding number must be an even number. 
The other topological number is the mirror Chern number which can be defined on a two-dimensional mirror plane in the Brillouin zone. We calculate the mirror Chern number as a function of the magnitude of the spin-orbit interaction and the pair potential using the method proposed by Suzuki $et$. $al$ \cite{doi:10.1143/JPSJ.74.1674}. In Fig.\ref{fig_phase}, we show the topological phase diagram of the STCI with $\Delta_2$. The color shows the band gap at the $L$ point. Here, we set $\mu=0.27$ eV. The mirror Chern number changes $-4 \rightarrow -2 \rightarrow 0$ with gap closing.  This is different from the case in the STI, since the mirror Chern number of the STI changes $-2 \rightarrow -1 \rightarrow 0$ \cite{PhysRevLett.108.107005}. 

For $\Delta_2$, the BdG Hamiltonian on the mirror invariant plane can be divided into two eigenvectors of the mirror operator in Eq.(\ref{mirror_BdG_m}), each of which belongs to class $D$. This state is the topological crystalline superconducting state defined in Ref. \cite{PhysRevLett.111.087002} since each subsector has particle-hole symmetry. 
\section{Surface spectral function}
\begin{figure}[tb]
\begin{center}
\includegraphics[width=8cm]{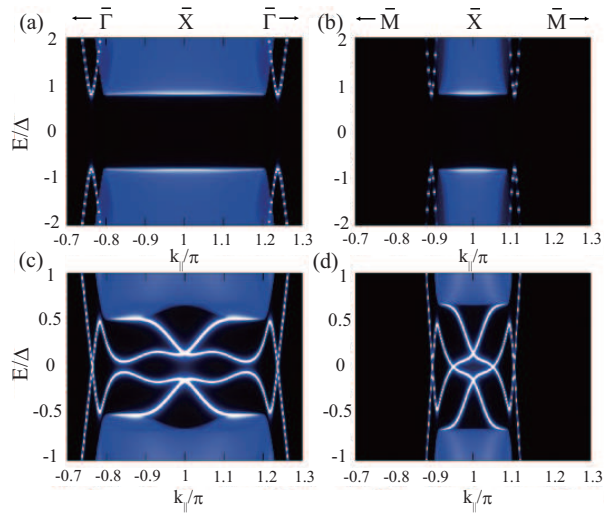}
\caption{Surface spectral function for the (001) surface. (a) and (b) show the spectral function for the even-parity spin-singlet pair potential $\Delta_1$ along mirror symmetric $\bar\Gamma$-$\bar X$-$\bar\Gamma$ line and $\bar M$-$\bar X$-$\bar M$ line, respectively. (c) and (d) show the spectral function for the odd-parity spin-triplet pair potential $\Delta_2$ along mirror symmetric $\bar\Gamma$-$\bar X$-$\bar\Gamma$ line and $\bar M$-$\bar X$-$\bar M$ line, respectively. In the case of $\Delta_1$, there is no Andreev bound state in the superconducting gap. On the other hand, there are mirror-protected zero-energy Andreev bound state.}
\label{fig_spe}
\end{center}
\end{figure}
\begin{figure}[b]
\begin{center}
\includegraphics[width=8.5cm]{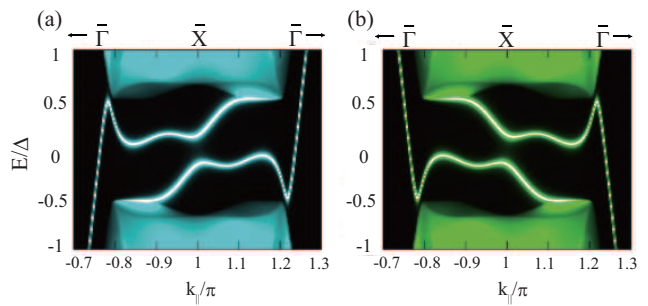}
\caption{Mirror separated spectral function in the case of $\Delta_2$ for (a) $+i$ and (b) $-i$ along the mirror symmetric line $\bar{\Gamma}$-$\bar{X}$-$\bar{\Gamma}$.}
\label{fig_mirror}
\end{center}
\end{figure}
\begin{figure*}[t]
\begin{center}
\includegraphics[width=15cm]{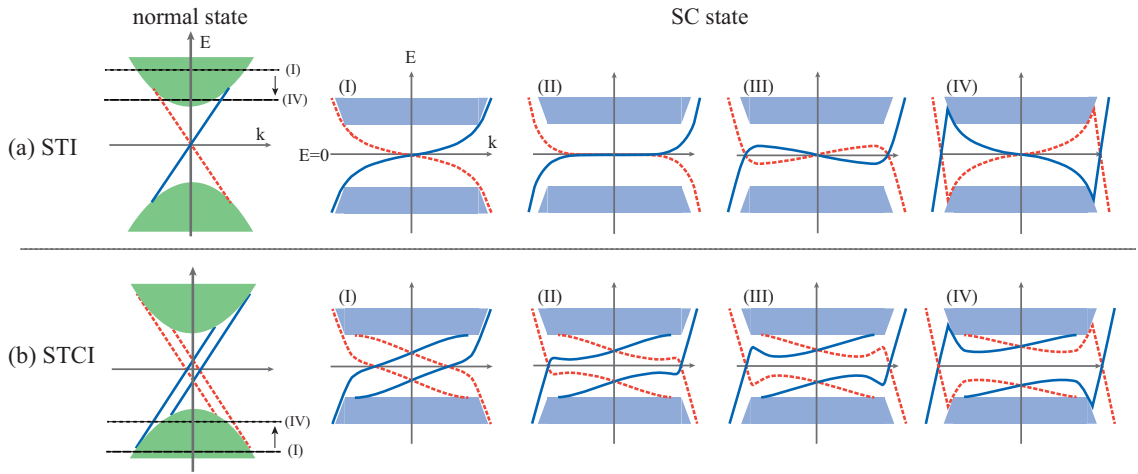}
\caption{Systematic change of the surface Andreev bound states (SABSs) in the superconducting topological insulator (STI) (a) and the superconducting topological crystalline insulator (STCI) (b). Blue solid (red dotted) lines show the SABSs for the mirror $+i$ ($-i$) sector. The SABSs change depending on the position of the chemical potential. Here, we assume Cu$_x$Bi$_2$Se$_3$ and Sn$_{1-x}$In$_{x}$Te for STI and STCI, respectively. For this reason, the figure of normal state for the STI (STCI) is depicted as the electron- (hole-) doped case. In both cases, carrier number decreases from the panel (I) to (IV).}
\label{fig_sche}
\end{center}
\end{figure*}
In this section, we present our numerically calculated results of the surface spectral function for the (001) surface. To obtain the surface spectral function, we use the recursive Green's function method using M$\rm \ddot o$bius transformation in the semi-finite system proposed by Umerski \cite{PhysRevB.55.5266}. See Supplemental Material for the details of the calculation method.
In this calculation, we set $\Delta=0.06$ eV in Eqs. (\ref{d1}) and (\ref{d2}). The chemical potential is taken as $\mu=-0.2$ eV to fit the size of Fermi surface observed in ARPES measurements \cite{PhysRevLett.110.206804}. In this case, the conduction band and the TCI surface state are well separated at the Fermi level, and the mirror Chern number is $-2$. 

First, we show the surface spectral function for $\Delta_1$. Figures. \ref{fig_spe} (a) and (b) show the spectral function along the $\bar\Gamma$-$\bar X$-$\bar\Gamma$ and $\bar M$-$\bar X$-$\bar M$ lines, respectively. In the case of $\Delta_1$, both bulk and TCI surface states have a superconducting gap and there is no inner gap state.
Next, in Figs. \ref{fig_spe} (c) and (d), we show the surface spectral function for $\Delta_2$. On the mirror symmetric line shown in Fig. \ref{fig_spe}(c), there are two zero-energy SABSs corresponding to the mirror Chern number $=-2$. 
\begin{figure}[b]
\begin{center}
\includegraphics[width=5.5cm]{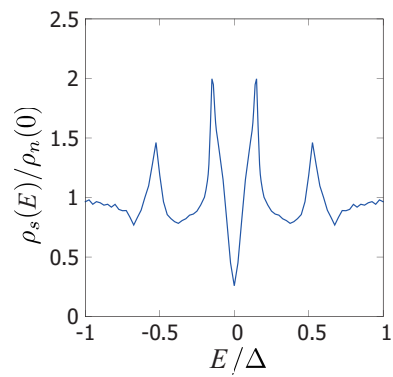}
\caption{Surface density of state in the case of $\Delta_2$ for the (001) surface. There are four peaks at $E/\Delta=\pm0.49$ and $\pm0.16$}
\label{fig_SDOS}
\end{center}
\end{figure}
Different from the simple linear dispersion in the TCI, the dispersion of the surface state in the superconducting state is twisted since the SABSs merge into the Dirac cone in the normal state which exist outside the Fermi wave vector. Using the mirror reflection operator, we decompose the surface spectral function into two subsectors with different eigenvalue $\pm i$ as shown in Fig. \ref{fig_mirror}.  As we mentioned in Sec. \ref{sec_topo}, the Chern number of each subsector $\pm i$ is $\mp 2$. Therefore, there are two chiral SABSs in each subsector. 
There are also zero-energy SABSs along $\bar M$-$\bar X$-$\bar M$ line [see Fig. \ref{fig_spe}(d)]. These zero energy SABSs can be interpreted by zero-dimensional topological number defined by combining with $\hat{{\cal M}}_{(110)}$ and $\hat{\cal C}$ \cite{2015arXiv150407408K}. These zero-energy SABSs disappear if we enlarge the magnitude of $\Delta$ like $\Delta=0.1$ eV.

We now turn to the detail analysis on the mirror protected SABSs which appears in the case of $\Delta_2$. 
First, we briefly review the systematic change of the time-reversal symmetry (TRS) protected SABSs in the STI such as Cu$_x$Bi$_2$Se$_3$ \cite{PhysRevB.83.134516,PhysRevLett.108.107005,PhysRevB.85.180509}. The dispersion of the SABSs in the STI changes mainly with the chemical potential. If the Fermi level is much higher than the energy where surface state is merged into bulk band, the shape of the SABSs is normal cone shape similar to that of the $^3$He BW phase, as can be seen in Fig. \ref{fig_sche} (a) (I). On the other hand, as lower the Fermi level, the SABSs start twisting as shown in Fig. \ref{fig_sche} (a) (III) and, finally, the SABSs become like Fig. \ref{fig_sche} (a) (IV). This is regarded as the Lifshitz transition of the surface state. At the transition point of this Lifshitz transition, the dispersion becomes flat like as shown in Fig. \ref{fig_sche} (a) (II) since the position of zero energy SABS is pinned at time reversal invariant momenta, i. e. $k=0$. 
This flat-like band can explain the ZBCP in the point-contact experiments even if the bulk energy spectrum is fully gapped \cite{PhysRevB.83.134516,PhysRevLett.108.107005,PhysRevB.85.180509}.
\begin{figure}[t]
\begin{center}
\includegraphics[width=8.5cm]{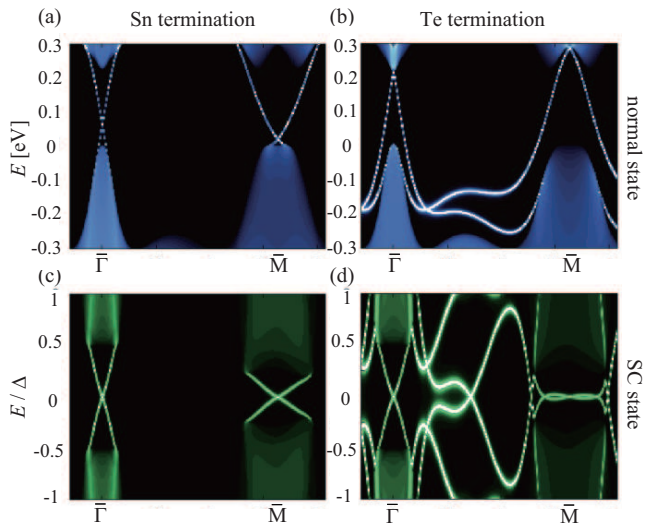}
\caption{Surface spectral function for the (111) surface along mirror-symmetric line $\bar \Gamma$-$\bar M$. (a) and (b) show the surface spectral function of the normal state for Sn- and Te-terminated surface, respectively. (c) and (d) show the spectral function of superconducting state for Sn- and Te-terminated surfaces, respectively. Here, we set the chemical potential $\mu=-0.2$ eV, which is the same as the calculation for the (001) surface. }
\label{fig_111}
\end{center}
\end{figure}

Next, we reveal the structural transition of the SABSs in the STCI. 
If the Fermi level is much lower than the energy where the TCI surface state is merged into the bulk band, there are two Dirac cones on the mirror plane as shown in Fig. \ref{fig_sche} (b) (I). On the other hand, in the same manner as the STI, if the chemical potential is located on the energy where surface and bulk states are well separated, the dispersion of the SABSs in the STCI is twisted as shown in Fig. \ref{fig_sche}(b)(IV). 
However, different from the SABSs in the STI, this change of the dispersion does not involve the Lifshitz transition. Thus, the surface state of the STCI does not have to host zero-energy flat dispersion between the twisting (I) and non-twisting (IV) cases as shown in Fig. \ref{fig_sche}.(b) (II) and (III). 
This can be understood by the difference of the symmetry which protects the zero energy SABSs. 
Different from the case of the STI, the zero-energy SABSs in the STCI are not protected by the TRS but the mirror symmetry. Therefore, the position of the zero-energy SABSs in the surface Brillouin zone can move along the mirror-symmetric line with doping. For this reason, the mirror-protected SABSs in the STCI can twist without presenting the flat-like band at zero energy.\\


\section{Interpretation of experiments and discussion}
\begin{figure}[t]
\begin{center}
\includegraphics[width=8cm]{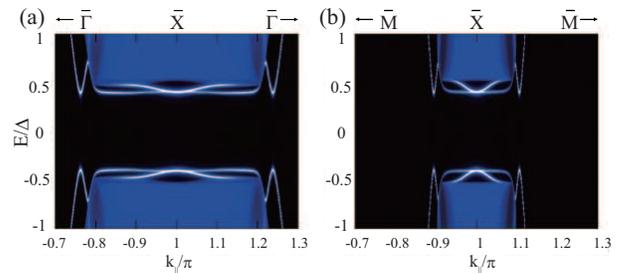}
\caption{Surface spectral function in the case of $\Delta_2+i\Delta_1$ pairing.  We introduce $\Delta_1$ and $\Delta_2$ at the same ratio. }
\label{fig_spe_p+is}
\end{center}
\end{figure}
We discuss the experimental situation for the tunneling spectroscopy. 
First, in Fig. \ref{fig_SDOS}, we show our numerically calculated results of normalized surface density of state $\rho_s(E)/\rho_n(0)$ for the (001) surface in the case of $\Delta_2$. Here, we set $\Delta=0.02$ eV. As can be seen from Fig.\ref{fig_SDOS}, there are four peaks at $E/\Delta=\pm0.49$ and $\pm0.16$. 
Therefore, it is reasonable that four peaks appear in conductance measurement with low transmissivity realized in scanning tunneling spectroscopy (STS).
On the other hand, in the point-contact measurements, it is likely to include the components of the surface with different orientations since the surface of crystal may contain many steps in the contact region. Thus, we calculate the surface spectral function for the (111) surface as an example of different surfaces. In the case of the (111) surface, $L$ points are projected to either $\bar \Gamma$ point or $\bar M$ points as shown in Fig. \ref{fig_cry}(b). Moreover, there are two distinct situations, i.e., Sn-terminated surface and Te-terminated surface. In Fig. \ref{fig_111} (a) [(b)], the spectral function for Sn- (Te-) terminated surface in the normal state is presented. In the case of the Sn-terminated surface, Dirac points are located on the high symmetric points $\bar{\Gamma}$ and $\bar{M}$. On the other hand, in the case of Te-terminated surface, Dirac points appear on the mirror symmetric line $\bar\Gamma$-$\bar M$. In Fig.\ref{fig_111} (c)((d)), we show the Sn(Te)-terminated surface spectral function in the superconducting state for $\Delta_2$. As can be seen from the figures, the SABSs for the (111) surface is completely different from those for the (001) surface. 
In the case of Sn-terminated surface, two cone-shape surface states appear at the $\bar \Gamma$ and $\bar M$ points on the surface Brillouin zone corresponding to mirror Chern number $n_M=-2$. In this case, the change of the dispersion of the SABS is almost the same as that of the STI. 
In the case of Te-terminated surface, as you can see in Fig.\ref{fig_111} (d), the surface state at $\bar M$ point is flat like and there are zero-energy surface states between the $\bar \Gamma$ and $\bar M$ points since the valley surface state remains due to the existence of the mirror reflection symmetry.

From the above calculation, it is reasonable to conclude that the bias voltage dependence of conductance for the (001) surface can be changed from the four peaks spectra if the component of the (111) surface is included.
In addition, in the normal state, it has been revealed that the edge state of (001) film dramatically changes depending on the number of layers \cite{PhysRevB.90.045309}. It can be expected that the similar situation may occur in the superconducting state due to the existence of the steps on the surface. 
Therefore, further calculations are necessary to interpret the obtained point-contact experiments \cite{PhysRevLett.109.217004}. 

Before closing this section, we briefly refer to the case of $\Delta_2+i\Delta_1$, i.e., $p+is$ pairing which breaks time-reversal symmetry.
Novak $et.$ $al.$ have indicated the realization of $p+is$ pairing in Sn$_{1-x}$In$_{x}$Te by point-contact experiments \cite{PhysRevB.88.140502}. Theoretically, Goswami and Roy have suggested the possibility of the $p+is$ pairing in doped topological materials with an effective model \cite{PhysRevB.90.041301}. Motivated by the studies, we calculate the spectral function for the $\Delta_2+i\Delta_1$ state in the lattice model. The calculated results are shown in Fig. \ref{fig_spe_p+is}. Consequently, we find that even in the lattice model, the system has gapped SABSs since $\Delta_2+i\Delta_1$ pairing breaks the (110) mirror symmetry. Our model is useful for further discussion of anomalous thermal Hall effect in Sn$_{1-x}$In$_{x}$Te.
\section{Conclusion}
In this paper, we have theoretically studied the SnTe class TCI in the superconducting state. We have introduced the possible fully-gapped pair potentials to the TCI and calculated the surface spectral function using the recursive Green's function method. 
We have found that the STCI hosts the mirror-protected SABSs when odd-parity pair potential $\Delta_2$ is realized. The mirror-protected SABSs for the (001) surface appear only when we use the model that describes the entire Brillouin zone. This result suggests the importance of considering the overlap of the Fermi surface to reveal the SABSs in general. We have also shown the systematic change of the dispersion of the mirror-protected SABSs. It has been revealed that the dispersion of the mirror-protected SABSs is not necessary to become flat at zero energy when it twists, which is different from the SABSs in the STI. 
Moreover, we find that four peaks appear in the SDOS, which can be detected by STS measurements. 

In analogy with the mirror-protected Dirac cone in the normal state, the mirror-protected SABSs can be tuned with the electronic field or strain. This means that some physical properties, e.g., the Josephson current can be controlled by these effects. We expect that further study of the mirror-protected SABS will lead to the development of superconducting electronics.
\section{Acknowledgements}\label{sec_Ack}
We thank A. Yamakage, S. Kobayashi and H. Ozawa for valuable discussions. 
This work was supported by the ``Topological Quantum Phenomena" Grant-in Aid for Scientific Research on Innovative Areas from the Ministry of Education, Culture, Sports, Science and Technology (MEXT) of Japan (No. 22103005), the ``Topological Materials Science" Grant-in Aid for Scientific Research on Innovative Areas from the MEXT of Japan (No. 15H05853, 15H05855), Grant-in-aid for JSPS Fellows (No. 26010542) (TH) and  Grant-in-Aid for Scientific Research B (No. 25287085) (MS).

\bibliography{ref}

\begin{thebibliography}{83}
\expandafter\ifx\csname natexlab\endcsname\relax\def\natexlab#1{#1}\fi
\expandafter\ifx\csname bibnamefont\endcsname\relax
  \def\bibnamefont#1{#1}\fi
\expandafter\ifx\csname bibfnamefont\endcsname\relax
  \def\bibfnamefont#1{#1}\fi
\expandafter\ifx\csname citenamefont\endcsname\relax
  \def\citenamefont#1{#1}\fi
\expandafter\ifx\csname url\endcsname\relax
  \def\url#1{\texttt{#1}}\fi
\expandafter\ifx\csname urlprefix\endcsname\relax\def\urlprefix{URL }\fi
\providecommand{\bibinfo}[2]{#2}
\providecommand{\eprint}[2][]{\url{#2}}

\bibitem[{\citenamefont{Kashiwaya and Tanaka}(2000)}]{ABS1}
\bibinfo{author}{\bibfnamefont{S.}~\bibnamefont{Kashiwaya}} \bibnamefont{and}
  \bibinfo{author}{\bibfnamefont{Y.}~\bibnamefont{Tanaka}},
  \bibinfo{journal}{Rep. Prog. Phys.} \textbf{\bibinfo{volume}{63}},
  \bibinfo{pages}{1641} (\bibinfo{year}{2000}).

\bibitem[{\citenamefont{L{\"o}fwander et~al.}(2001)\citenamefont{L{\"o}fwander,
  Shumeiko, and Wendin}}]{ABS2}
\bibinfo{author}{\bibfnamefont{T.}~\bibnamefont{L{\"o}fwander}},
  \bibinfo{author}{\bibfnamefont{V.~S.} \bibnamefont{Shumeiko}},
  \bibnamefont{and} \bibinfo{author}{\bibfnamefont{G.}~\bibnamefont{Wendin}},
  \bibinfo{journal}{Supercond. Sci. Technol.} \textbf{\bibinfo{volume}{14}},
  \bibinfo{pages}{R53} (\bibinfo{year}{2001}).

\bibitem[{\citenamefont{Deutscher}(2005)}]{Deutscher}
\bibinfo{author}{\bibfnamefont{G.}~\bibnamefont{Deutscher}},
  \bibinfo{journal}{Rev. Mod. Phys.} \textbf{\bibinfo{volume}{77}},
  \bibinfo{pages}{109} (\bibinfo{year}{2005}).

\bibitem[{\citenamefont{Buchholtz and Zwicknagl}(1981)}]{ABS}
\bibinfo{author}{\bibfnamefont{L.~J.} \bibnamefont{Buchholtz}}
  \bibnamefont{and}
  \bibinfo{author}{\bibfnamefont{G.}~\bibnamefont{Zwicknagl}},
  \bibinfo{journal}{Phys. Rev. B} \textbf{\bibinfo{volume}{23}},
  \bibinfo{pages}{5788} (\bibinfo{year}{1981}).

\bibitem[{\citenamefont{Hara and Nagai}(1986)}]{ABSb}
\bibinfo{author}{\bibfnamefont{J.}~\bibnamefont{Hara}} \bibnamefont{and}
  \bibinfo{author}{\bibfnamefont{K.}~\bibnamefont{Nagai}},
  \bibinfo{journal}{Prog. Theor. Phys.} \textbf{\bibinfo{volume}{76}},
  \bibinfo{pages}{1237} (\bibinfo{year}{1986}).

\bibitem[{\citenamefont{Hu}(1994)}]{Hu}
\bibinfo{author}{\bibfnamefont{C.~R.} \bibnamefont{Hu}},
  \bibinfo{journal}{Phys. Rev. Lett.} \textbf{\bibinfo{volume}{72}},
  \bibinfo{pages}{1526} (\bibinfo{year}{1994}).

\bibitem[{\citenamefont{Tanaka and Kashiwaya}(1995)}]{TK95}
\bibinfo{author}{\bibfnamefont{Y.}~\bibnamefont{Tanaka}} \bibnamefont{and}
  \bibinfo{author}{\bibfnamefont{S.}~\bibnamefont{Kashiwaya}},
  \bibinfo{journal}{Phys. Rev. Lett.} \textbf{\bibinfo{volume}{74}},
  \bibinfo{pages}{3451} (\bibinfo{year}{1995}).

\bibitem[{\citenamefont{Yamashiro et~al.}(1997)\citenamefont{Yamashiro, Tanaka,
  and Kashiwaya}}]{Yamashiro}
\bibinfo{author}{\bibfnamefont{M.}~\bibnamefont{Yamashiro}},
  \bibinfo{author}{\bibfnamefont{Y.}~\bibnamefont{Tanaka}}, \bibnamefont{and}
  \bibinfo{author}{\bibfnamefont{S.}~\bibnamefont{Kashiwaya}},
  \bibinfo{journal}{Phys. Rev. B} \textbf{\bibinfo{volume}{56}},
  \bibinfo{pages}{7847} (\bibinfo{year}{1997}).

\bibitem[{\citenamefont{Kashiwaya et~al.}(2011)\citenamefont{Kashiwaya,
  Kashiwaya, Kambara, Furuta, Yaguchi, Tanaka, and Maeno}}]{Kashiwaya11}
\bibinfo{author}{\bibfnamefont{S.}~\bibnamefont{Kashiwaya}},
  \bibinfo{author}{\bibfnamefont{H.}~\bibnamefont{Kashiwaya}},
  \bibinfo{author}{\bibfnamefont{H.}~\bibnamefont{Kambara}},
  \bibinfo{author}{\bibfnamefont{T.}~\bibnamefont{Furuta}},
  \bibinfo{author}{\bibfnamefont{H.}~\bibnamefont{Yaguchi}},
  \bibinfo{author}{\bibfnamefont{Y.}~\bibnamefont{Tanaka}}, \bibnamefont{and}
  \bibinfo{author}{\bibfnamefont{Y.}~\bibnamefont{Maeno}},
  \bibinfo{journal}{Phys. Rev. Lett.} \textbf{\bibinfo{volume}{107}},
  \bibinfo{pages}{077003} (\bibinfo{year}{2011}).

\bibitem[{\citenamefont{Tanaka et~al.}(2012{\natexlab{a}})\citenamefont{Tanaka,
  Sato, and Nagaosa}}]{tanaka12}
\bibinfo{author}{\bibfnamefont{Y.}~\bibnamefont{Tanaka}},
  \bibinfo{author}{\bibfnamefont{M.}~\bibnamefont{Sato}}, \bibnamefont{and}
  \bibinfo{author}{\bibfnamefont{N.}~\bibnamefont{Nagaosa}},
  \bibinfo{journal}{J. Phys. Soc. Jpn.} \textbf{\bibinfo{volume}{81}},
  \bibinfo{pages}{011013} (\bibinfo{year}{2012}{\natexlab{a}}).

\bibitem[{\citenamefont{Alicea}(2012)}]{alicea12}
\bibinfo{author}{\bibfnamefont{J.}~\bibnamefont{Alicea}},
  \bibinfo{journal}{Rep. Prog. Phys.} \textbf{\bibinfo{volume}{75}},
  \bibinfo{pages}{076501} (\bibinfo{year}{2012}).

\bibitem[{\citenamefont{Wong et~al.}(2013)\citenamefont{Wong, Liu, Law, and
  Lee}}]{Law}
\bibinfo{author}{\bibfnamefont{C.~L.~M.} \bibnamefont{Wong}},
  \bibinfo{author}{\bibfnamefont{J.}~\bibnamefont{Liu}},
  \bibinfo{author}{\bibfnamefont{K.~T.} \bibnamefont{Law}}, \bibnamefont{and}
  \bibinfo{author}{\bibfnamefont{P.~A.} \bibnamefont{Lee}},
  \bibinfo{journal}{Phys. Rev. B} \textbf{\bibinfo{volume}{88}},
  \bibinfo{pages}{060504} (\bibinfo{year}{2013}).

\bibitem[{\citenamefont{Brydon et~al.}(2011)\citenamefont{Brydon, Schnyder, and
  Timm}}]{Brydon1}
\bibinfo{author}{\bibfnamefont{P.~M.~R.} \bibnamefont{Brydon}},
  \bibinfo{author}{\bibfnamefont{A.~P.} \bibnamefont{Schnyder}},
  \bibnamefont{and} \bibinfo{author}{\bibfnamefont{C.}~\bibnamefont{Timm}},
  \bibinfo{journal}{Phys. Rev. B} \textbf{\bibinfo{volume}{84}},
  \bibinfo{pages}{020501} (\bibinfo{year}{2011}).

\bibitem[{\citenamefont{Schnyder and Ryu}(2011)}]{Schnyder2011}
\bibinfo{author}{\bibfnamefont{A.~P.} \bibnamefont{Schnyder}} \bibnamefont{and}
  \bibinfo{author}{\bibfnamefont{S.}~\bibnamefont{Ryu}},
  \bibinfo{journal}{Phys. Rev. B} \textbf{\bibinfo{volume}{84}},
  \bibinfo{pages}{060504} (\bibinfo{year}{2011}).

\bibitem[{\citenamefont{Schnyder and Brydon}(2015)}]{SchnyderBrydon}
\bibinfo{author}{\bibfnamefont{A.~P.} \bibnamefont{Schnyder}} \bibnamefont{and}
  \bibinfo{author}{\bibfnamefont{P.~M.~R.} \bibnamefont{Brydon}},
  \bibinfo{journal}{J. Phys.: Condens. Matter} \textbf{\bibinfo{volume}{27}},
  \bibinfo{pages}{243201} (\bibinfo{year}{2015}).

\bibitem[{\citenamefont{Sato et~al.}(2011)\citenamefont{Sato, Tanaka, Yada, and
  Yokoyama}}]{STYY11}
\bibinfo{author}{\bibfnamefont{M.}~\bibnamefont{Sato}},
  \bibinfo{author}{\bibfnamefont{Y.}~\bibnamefont{Tanaka}},
  \bibinfo{author}{\bibfnamefont{K.}~\bibnamefont{Yada}}, \bibnamefont{and}
  \bibinfo{author}{\bibfnamefont{T.}~\bibnamefont{Yokoyama}},
  \bibinfo{journal}{Phys.\ Rev.\ B} \textbf{\bibinfo{volume}{83}},
  \bibinfo{pages}{224511} (\bibinfo{year}{2011}).

\bibitem[{\citenamefont{Tanaka et~al.}(2010)\citenamefont{Tanaka, Mizuno,
  Yokoyama, Yada, and Sato}}]{TMYYS10}
\bibinfo{author}{\bibfnamefont{Y.}~\bibnamefont{Tanaka}},
  \bibinfo{author}{\bibfnamefont{Y.}~\bibnamefont{Mizuno}},
  \bibinfo{author}{\bibfnamefont{T.}~\bibnamefont{Yokoyama}},
  \bibinfo{author}{\bibfnamefont{K.}~\bibnamefont{Yada}}, \bibnamefont{and}
  \bibinfo{author}{\bibfnamefont{M.}~\bibnamefont{Sato}},
  \bibinfo{journal}{Phys. Rev. Lett.} \textbf{\bibinfo{volume}{105}},
  \bibinfo{pages}{097002} (\bibinfo{year}{2010}).

\bibitem[{\citenamefont{Yada et~al.}(2011)\citenamefont{Yada, Sato, Tanaka, and
  Yokoyama}}]{YSTY10}
\bibinfo{author}{\bibfnamefont{K.}~\bibnamefont{Yada}},
  \bibinfo{author}{\bibfnamefont{M.}~\bibnamefont{Sato}},
  \bibinfo{author}{\bibfnamefont{Y.}~\bibnamefont{Tanaka}}, \bibnamefont{and}
  \bibinfo{author}{\bibfnamefont{T.}~\bibnamefont{Yokoyama}},
  \bibinfo{journal}{Phys. Rev. B.} \textbf{\bibinfo{volume}{83}},
  \bibinfo{pages}{064505} (\bibinfo{year}{2011}).

\bibitem[{\citenamefont{Sato et~al.}(2009)\citenamefont{Sato, Takahashi, and
  Fujimoto}}]{STF09}
\bibinfo{author}{\bibfnamefont{M.}~\bibnamefont{Sato}},
  \bibinfo{author}{\bibfnamefont{Y.}~\bibnamefont{Takahashi}},
  \bibnamefont{and} \bibinfo{author}{\bibfnamefont{S.}~\bibnamefont{Fujimoto}},
  \bibinfo{journal}{Phys. Rev. Lett.} \textbf{\bibinfo{volume}{103}},
  \bibinfo{pages}{020401} (\bibinfo{year}{2009}).

\bibitem[{\citenamefont{Lutchyn et~al.}(2010)\citenamefont{Lutchyn, Sau, and
  Das~Sarma}}]{lutchyn10}
\bibinfo{author}{\bibfnamefont{R.~M.} \bibnamefont{Lutchyn}},
  \bibinfo{author}{\bibfnamefont{J.~D.} \bibnamefont{Sau}}, \bibnamefont{and}
  \bibinfo{author}{\bibfnamefont{S.}~\bibnamefont{Das~Sarma}},
  \bibinfo{journal}{Phys. Rev. Lett.} \textbf{\bibinfo{volume}{105}},
  \bibinfo{pages}{077001} (\bibinfo{year}{2010}).

\bibitem[{\citenamefont{Oreg et~al.}(2010)\citenamefont{Oreg, Refael, and von
  Oppen}}]{oreg10}
\bibinfo{author}{\bibfnamefont{Y.}~\bibnamefont{Oreg}},
  \bibinfo{author}{\bibfnamefont{G.}~\bibnamefont{Refael}}, \bibnamefont{and}
  \bibinfo{author}{\bibfnamefont{F.}~\bibnamefont{von Oppen}},
  \bibinfo{journal}{Phys. Rev. Lett.} \textbf{\bibinfo{volume}{105}},
  \bibinfo{pages}{177002} (\bibinfo{year}{2010}).

\bibitem[{\citenamefont{Alicea}(2010)}]{alicea10}
\bibinfo{author}{\bibfnamefont{J.}~\bibnamefont{Alicea}},
  \bibinfo{journal}{Phys.\ Rev.\ B} \textbf{\bibinfo{volume}{81}},
  \bibinfo{pages}{125318} (\bibinfo{year}{2010}).

\bibitem[{\citenamefont{Hasan and Kane}(2010)}]{RevModPhys.82.3045}
\bibinfo{author}{\bibfnamefont{M.~Z.} \bibnamefont{Hasan}} \bibnamefont{and}
  \bibinfo{author}{\bibfnamefont{C.~L.} \bibnamefont{Kane}},
  \bibinfo{journal}{Rev. Mod. Phys.} \textbf{\bibinfo{volume}{82}},
  \bibinfo{pages}{3045} (\bibinfo{year}{2010}),
  \urlprefix\url{http://link.aps.org/doi/10.1103/RevModPhys.82.3045}.

\bibitem[{\citenamefont{Qi and Zhang}(2011)}]{RevModPhys.83.1057}
\bibinfo{author}{\bibfnamefont{X.-L.} \bibnamefont{Qi}} \bibnamefont{and}
  \bibinfo{author}{\bibfnamefont{S.-C.} \bibnamefont{Zhang}},
  \bibinfo{journal}{Rev. Mod. Phys.} \textbf{\bibinfo{volume}{83}},
  \bibinfo{pages}{1057} (\bibinfo{year}{2011}),
  \urlprefix\url{http://link.aps.org/doi/10.1103/RevModPhys.83.1057}.

\bibitem[{\citenamefont{Ando}(2013)}]{doi:10.7566/JPSJ.82.102001}
\bibinfo{author}{\bibfnamefont{Y.}~\bibnamefont{Ando}},
  \bibinfo{journal}{Journal of the Physical Society of Japan}
  \textbf{\bibinfo{volume}{82}}, \bibinfo{pages}{102001}
  (\bibinfo{year}{2013}).

\bibitem[{\citenamefont{Sasaki et~al.}(2011)\citenamefont{Sasaki, Kriener,
  Segawa, Yada, Tanaka, Sato, and Ando}}]{Sasaki}
\bibinfo{author}{\bibfnamefont{S.}~\bibnamefont{Sasaki}},
  \bibinfo{author}{\bibfnamefont{M.}~\bibnamefont{Kriener}},
  \bibinfo{author}{\bibfnamefont{K.}~\bibnamefont{Segawa}},
  \bibinfo{author}{\bibfnamefont{K.}~\bibnamefont{Yada}},
  \bibinfo{author}{\bibfnamefont{Y.}~\bibnamefont{Tanaka}},
  \bibinfo{author}{\bibfnamefont{M.}~\bibnamefont{Sato}}, \bibnamefont{and}
  \bibinfo{author}{\bibfnamefont{Y.}~\bibnamefont{Ando}},
  \bibinfo{journal}{Phys. Rev. Lett.} \textbf{\bibinfo{volume}{107}},
  \bibinfo{pages}{217001} (\bibinfo{year}{2011}).

\bibitem[{\citenamefont{Hor et~al.}(2010)\citenamefont{Hor, Williams,
  Checkelsky, Roushan, Seo, Xu, Zandbergen, Yazdani, Ong, and Cava}}]{Hor}
\bibinfo{author}{\bibfnamefont{Y.~S.} \bibnamefont{Hor}},
  \bibinfo{author}{\bibfnamefont{A.~J.} \bibnamefont{Williams}},
  \bibinfo{author}{\bibfnamefont{J.~G.} \bibnamefont{Checkelsky}},
  \bibinfo{author}{\bibfnamefont{P.}~\bibnamefont{Roushan}},
  \bibinfo{author}{\bibfnamefont{J.}~\bibnamefont{Seo}},
  \bibinfo{author}{\bibfnamefont{Q.}~\bibnamefont{Xu}},
  \bibinfo{author}{\bibfnamefont{H.~W.} \bibnamefont{Zandbergen}},
  \bibinfo{author}{\bibfnamefont{A.}~\bibnamefont{Yazdani}},
  \bibinfo{author}{\bibfnamefont{N.~P.} \bibnamefont{Ong}}, \bibnamefont{and}
  \bibinfo{author}{\bibfnamefont{R.~J.} \bibnamefont{Cava}},
  \bibinfo{journal}{Phys. Rev. Lett.} \textbf{\bibinfo{volume}{104}},
  \bibinfo{pages}{057001} (\bibinfo{year}{2010}).

\bibitem[{\citenamefont{Sasaki and Mizushima}(2015)}]{Sasaki2015206}
\bibinfo{author}{\bibfnamefont{S.}~\bibnamefont{Sasaki}} \bibnamefont{and}
  \bibinfo{author}{\bibfnamefont{T.}~\bibnamefont{Mizushima}},
  \bibinfo{journal}{Physica C: Superconductivity and its Applications}
  \textbf{\bibinfo{volume}{514}}, \bibinfo{pages}{206 } (\bibinfo{year}{2015}),
  ISSN \bibinfo{issn}{0921-4534}, \bibinfo{note}{superconducting Materials:
  Conventional, Unconventional and Undetermined}.

\bibitem[{\citenamefont{Hashimoto et~al.}(2014)\citenamefont{Hashimoto, Yada,
  Yamakage, Sato, and Tanaka}}]{0953-2048-27-10-104002}
\bibinfo{author}{\bibfnamefont{T.}~\bibnamefont{Hashimoto}},
  \bibinfo{author}{\bibfnamefont{K.}~\bibnamefont{Yada}},
  \bibinfo{author}{\bibfnamefont{A.}~\bibnamefont{Yamakage}},
  \bibinfo{author}{\bibfnamefont{M.}~\bibnamefont{Sato}}, \bibnamefont{and}
  \bibinfo{author}{\bibfnamefont{Y.}~\bibnamefont{Tanaka}},
  \bibinfo{journal}{Superconductor Science and Technology}
  \textbf{\bibinfo{volume}{27}}, \bibinfo{pages}{104002}
  (\bibinfo{year}{2014}).

\bibitem[{\citenamefont{Hashimoto et~al.}(2013)\citenamefont{Hashimoto, Yada,
  Yamakage, Sato, and Tanaka}}]{doi:10.7566/JPSJ.82.044704}
\bibinfo{author}{\bibfnamefont{T.}~\bibnamefont{Hashimoto}},
  \bibinfo{author}{\bibfnamefont{K.}~\bibnamefont{Yada}},
  \bibinfo{author}{\bibfnamefont{A.}~\bibnamefont{Yamakage}},
  \bibinfo{author}{\bibfnamefont{M.}~\bibnamefont{Sato}}, \bibnamefont{and}
  \bibinfo{author}{\bibfnamefont{Y.}~\bibnamefont{Tanaka}},
  \bibinfo{journal}{Journal of the Physical Society of Japan}
  \textbf{\bibinfo{volume}{82}}, \bibinfo{pages}{044704}
  (\bibinfo{year}{2013}).

\bibitem[{\citenamefont{Takami et~al.}(2014)\citenamefont{Takami, Yada,
  Yamakage, Sato, and Tanaka}}]{doi:10.7566/JPSJ.83.064705}
\bibinfo{author}{\bibfnamefont{S.}~\bibnamefont{Takami}},
  \bibinfo{author}{\bibfnamefont{K.}~\bibnamefont{Yada}},
  \bibinfo{author}{\bibfnamefont{A.}~\bibnamefont{Yamakage}},
  \bibinfo{author}{\bibfnamefont{M.}~\bibnamefont{Sato}}, \bibnamefont{and}
  \bibinfo{author}{\bibfnamefont{Y.}~\bibnamefont{Tanaka}},
  \bibinfo{journal}{Journal of the Physical Society of Japan}
  \textbf{\bibinfo{volume}{83}}, \bibinfo{pages}{064705}
  (\bibinfo{year}{2014}).

\bibitem[{\citenamefont{Nagai et~al.}(2012)\citenamefont{Nagai, Nakamura, and
  Machida}}]{Nagai1}
\bibinfo{author}{\bibfnamefont{Y.}~\bibnamefont{Nagai}},
  \bibinfo{author}{\bibfnamefont{H.}~\bibnamefont{Nakamura}}, \bibnamefont{and}
  \bibinfo{author}{\bibfnamefont{M.}~\bibnamefont{Machida}},
  \bibinfo{journal}{Phys. Rev. B} \textbf{\bibinfo{volume}{86}},
  \bibinfo{pages}{094507} (\bibinfo{year}{2012}).

\bibitem[{\citenamefont{Nagai et~al.}(2015)\citenamefont{Nagai, Nakamura, and
  Machida}}]{Nagai2}
\bibinfo{author}{\bibfnamefont{Y.}~\bibnamefont{Nagai}},
  \bibinfo{author}{\bibfnamefont{H.}~\bibnamefont{Nakamura}}, \bibnamefont{and}
  \bibinfo{author}{\bibfnamefont{M.}~\bibnamefont{Machida}},
  \bibinfo{journal}{Journal of the Physical Society of Japan}
  \textbf{\bibinfo{volume}{84}}, \bibinfo{pages}{033703}
  (\bibinfo{year}{2015}).

\bibitem[{\citenamefont{Nagai}(2014)}]{Nagai3}
\bibinfo{author}{\bibfnamefont{Y.}~\bibnamefont{Nagai}},
  \bibinfo{journal}{Journal of the Physical Society of Japan}
  \textbf{\bibinfo{volume}{83}}, \bibinfo{pages}{063705}
  (\bibinfo{year}{2014}).

\bibitem[{\citenamefont{Nagai et~al.}(2014)\citenamefont{Nagai, Nakamura, and
  Machida}}]{Nagai4}
\bibinfo{author}{\bibfnamefont{Y.}~\bibnamefont{Nagai}},
  \bibinfo{author}{\bibfnamefont{H.}~\bibnamefont{Nakamura}}, \bibnamefont{and}
  \bibinfo{author}{\bibfnamefont{M.}~\bibnamefont{Machida}},
  \bibinfo{journal}{Journal of the Physical Society of Japan}
  \textbf{\bibinfo{volume}{83}}, \bibinfo{pages}{064703}
  (\bibinfo{year}{2014}).

\bibitem[{\citenamefont{Yip}(2013)}]{Yip}
\bibinfo{author}{\bibfnamefont{S.-K.} \bibnamefont{Yip}},
  \bibinfo{journal}{Phys. Rev. B} \textbf{\bibinfo{volume}{87}},
  \bibinfo{pages}{104505} (\bibinfo{year}{2013}).

\bibitem[{\citenamefont{Fu and Berg}(2010)}]{FuBerg}
\bibinfo{author}{\bibfnamefont{L.}~\bibnamefont{Fu}} \bibnamefont{and}
  \bibinfo{author}{\bibfnamefont{E.}~\bibnamefont{Berg}},
  \bibinfo{journal}{Phys. Rev. Lett.} \textbf{\bibinfo{volume}{105}},
  \bibinfo{pages}{097001} (\bibinfo{year}{2010}).

\bibitem[{\citenamefont{Fu}(2014)}]{PhysRevB.90.100509}
\bibinfo{author}{\bibfnamefont{L.}~\bibnamefont{Fu}}, \bibinfo{journal}{Phys.
  Rev. B} \textbf{\bibinfo{volume}{90}}, \bibinfo{pages}{100509}
  (\bibinfo{year}{2014}).

\bibitem[{\citenamefont{Kriener et~al.}(2011)\citenamefont{Kriener, Segawa,
  Ren, Sasaki, and Ando}}]{Kriener1}
\bibinfo{author}{\bibfnamefont{M.}~\bibnamefont{Kriener}},
  \bibinfo{author}{\bibfnamefont{K.}~\bibnamefont{Segawa}},
  \bibinfo{author}{\bibfnamefont{Z.}~\bibnamefont{Ren}},
  \bibinfo{author}{\bibfnamefont{S.}~\bibnamefont{Sasaki}}, \bibnamefont{and}
  \bibinfo{author}{\bibfnamefont{Y.}~\bibnamefont{Ando}},
  \bibinfo{journal}{Phys. Rev. Lett.} \textbf{\bibinfo{volume}{106}},
  \bibinfo{pages}{127004} (\bibinfo{year}{2011}).

\bibitem[{\citenamefont{Kriener et~al.}(2012)\citenamefont{Kriener, Segawa,
  Sasaki, and Ando}}]{Kriener2}
\bibinfo{author}{\bibfnamefont{M.}~\bibnamefont{Kriener}},
  \bibinfo{author}{\bibfnamefont{K.}~\bibnamefont{Segawa}},
  \bibinfo{author}{\bibfnamefont{S.}~\bibnamefont{Sasaki}}, \bibnamefont{and}
  \bibinfo{author}{\bibfnamefont{Y.}~\bibnamefont{Ando}},
  \bibinfo{journal}{Phys. Rev. B} \textbf{\bibinfo{volume}{86}},
  \bibinfo{pages}{180505} (\bibinfo{year}{2012}).

\bibitem[{\citenamefont{Bay et~al.}(2012)\citenamefont{Bay, Naka, Huang,
  Luigjes, Golden, and de~Visser}}]{Bay}
\bibinfo{author}{\bibfnamefont{T.~V.} \bibnamefont{Bay}},
  \bibinfo{author}{\bibfnamefont{T.}~\bibnamefont{Naka}},
  \bibinfo{author}{\bibfnamefont{Y.~K.} \bibnamefont{Huang}},
  \bibinfo{author}{\bibfnamefont{H.}~\bibnamefont{Luigjes}},
  \bibinfo{author}{\bibfnamefont{M.~S.} \bibnamefont{Golden}},
  \bibnamefont{and}
  \bibinfo{author}{\bibfnamefont{A.}~\bibnamefont{de~Visser}},
  \bibinfo{journal}{Phys. Rev. Lett.} \textbf{\bibinfo{volume}{108}},
  \bibinfo{pages}{057001} (\bibinfo{year}{2012}).

\bibitem[{\citenamefont{Zocher and Rosenow}(2013)}]{Zocher}
\bibinfo{author}{\bibfnamefont{B.}~\bibnamefont{Zocher}} \bibnamefont{and}
  \bibinfo{author}{\bibfnamefont{B.}~\bibnamefont{Rosenow}},
  \bibinfo{journal}{Phys. Rev. B} \textbf{\bibinfo{volume}{87}},
  \bibinfo{pages}{155138} (\bibinfo{year}{2013}).

\bibitem[{\citenamefont{Brydon et~al.}(2014)\citenamefont{Brydon, Das~Sarma,
  Hui, and Sau}}]{PhysRevB.90.184512}
\bibinfo{author}{\bibfnamefont{P.~M.~R.} \bibnamefont{Brydon}},
  \bibinfo{author}{\bibfnamefont{S.}~\bibnamefont{Das~Sarma}},
  \bibinfo{author}{\bibfnamefont{H.-Y.} \bibnamefont{Hui}}, \bibnamefont{and}
  \bibinfo{author}{\bibfnamefont{J.~D.} \bibnamefont{Sau}},
  \bibinfo{journal}{Phys. Rev. B} \textbf{\bibinfo{volume}{90}},
  \bibinfo{pages}{184512} (\bibinfo{year}{2014}).

\bibitem[{\citenamefont{Yamakage et~al.}(2012)\citenamefont{Yamakage, Yada,
  Sato, and Tanaka}}]{PhysRevB.85.180509}
\bibinfo{author}{\bibfnamefont{A.}~\bibnamefont{Yamakage}},
  \bibinfo{author}{\bibfnamefont{K.}~\bibnamefont{Yada}},
  \bibinfo{author}{\bibfnamefont{M.}~\bibnamefont{Sato}}, \bibnamefont{and}
  \bibinfo{author}{\bibfnamefont{Y.}~\bibnamefont{Tanaka}},
  \bibinfo{journal}{Phys. Rev. B} \textbf{\bibinfo{volume}{85}},
  \bibinfo{pages}{180509} (\bibinfo{year}{2012}).

\bibitem[{\citenamefont{Hsieh and Fu}(2012)}]{PhysRevLett.108.107005}
\bibinfo{author}{\bibfnamefont{T.~H.} \bibnamefont{Hsieh}} \bibnamefont{and}
  \bibinfo{author}{\bibfnamefont{L.}~\bibnamefont{Fu}}, \bibinfo{journal}{Phys.
  Rev. Lett.} \textbf{\bibinfo{volume}{108}}, \bibinfo{pages}{107005}
  (\bibinfo{year}{2012}).

\bibitem[{\citenamefont{Hao and Lee}(2011)}]{PhysRevB.83.134516}
\bibinfo{author}{\bibfnamefont{L.}~\bibnamefont{Hao}} \bibnamefont{and}
  \bibinfo{author}{\bibfnamefont{T.~K.} \bibnamefont{Lee}},
  \bibinfo{journal}{Phys. Rev. B} \textbf{\bibinfo{volume}{83}},
  \bibinfo{pages}{134516} (\bibinfo{year}{2011}).

\bibitem[{\citenamefont{Kobayashi and Sato}(2015)}]{2015arXiv150407408K}
\bibinfo{author}{\bibfnamefont{S.}~\bibnamefont{Kobayashi}} \bibnamefont{and}
  \bibinfo{author}{\bibfnamefont{M.}~\bibnamefont{Sato}},
  \bibinfo{journal}{Phys. Rev. Lett.} \textbf{\bibinfo{volume}{115}},
  \bibinfo{pages}{187001} (\bibinfo{year}{2015}).

\bibitem[{\citenamefont{Lu et~al.}(2015)\citenamefont{Lu, Yada, Sato, and
  Tanaka}}]{PhysRevLett.114.096804}
\bibinfo{author}{\bibfnamefont{B.}~\bibnamefont{Lu}},
  \bibinfo{author}{\bibfnamefont{K.}~\bibnamefont{Yada}},
  \bibinfo{author}{\bibfnamefont{M.}~\bibnamefont{Sato}}, \bibnamefont{and}
  \bibinfo{author}{\bibfnamefont{Y.}~\bibnamefont{Tanaka}},
  \bibinfo{journal}{Phys. Rev. Lett.} \textbf{\bibinfo{volume}{114}},
  \bibinfo{pages}{096804} (\bibinfo{year}{2015}).

\bibitem[{\citenamefont{Bednik et~al.}(2015)\citenamefont{Bednik, Zyuzin, and
  Burkov}}]{PhysRevB.92.035153}
\bibinfo{author}{\bibfnamefont{G.}~\bibnamefont{Bednik}},
  \bibinfo{author}{\bibfnamefont{A.~A.} \bibnamefont{Zyuzin}},
  \bibnamefont{and} \bibinfo{author}{\bibfnamefont{A.~A.}
  \bibnamefont{Burkov}}, \bibinfo{journal}{Phys. Rev. B}
  \textbf{\bibinfo{volume}{92}}, \bibinfo{pages}{035153}
  (\bibinfo{year}{2015}).

\bibitem[{\citenamefont{Fu}(2011)}]{PhysRevLett.106.106802}
\bibinfo{author}{\bibfnamefont{L.}~\bibnamefont{Fu}}, \bibinfo{journal}{Phys.
  Rev. Lett.} \textbf{\bibinfo{volume}{106}}, \bibinfo{pages}{106802}
  (\bibinfo{year}{2011}).

\bibitem[{\citenamefont{Hsieh et~al.}(2012)\citenamefont{Hsieh, Lin, Liu, Duan,
  Bansil, and Fu}}]{NatCommFu}
\bibinfo{author}{\bibfnamefont{T.~H.} \bibnamefont{Hsieh}},
  \bibinfo{author}{\bibfnamefont{H.}~\bibnamefont{Lin}},
  \bibinfo{author}{\bibfnamefont{J.}~\bibnamefont{Liu}},
  \bibinfo{author}{\bibfnamefont{W.}~\bibnamefont{Duan}},
  \bibinfo{author}{\bibfnamefont{A.}~\bibnamefont{Bansil}}, \bibnamefont{and}
  \bibinfo{author}{\bibfnamefont{L.}~\bibnamefont{Fu}},
  \bibinfo{journal}{Nature Communications} \textbf{\bibinfo{volume}{3}},
  \bibinfo{pages}{982} (\bibinfo{year}{2012}).

\bibitem[{\citenamefont{Tanaka et~al.}(2012{\natexlab{b}})\citenamefont{Tanaka,
  Ren, Sato, Nakayama, Souma, Takahashi, Segawa, and Ando}}]{SnTe_ARPES}
\bibinfo{author}{\bibfnamefont{Y.}~\bibnamefont{Tanaka}},
  \bibinfo{author}{\bibfnamefont{Z.}~\bibnamefont{Ren}},
  \bibinfo{author}{\bibfnamefont{T.}~\bibnamefont{Sato}},
  \bibinfo{author}{\bibfnamefont{K.}~\bibnamefont{Nakayama}},
  \bibinfo{author}{\bibfnamefont{S.}~\bibnamefont{Souma}},
  \bibinfo{author}{\bibfnamefont{T.}~\bibnamefont{Takahashi}},
  \bibinfo{author}{\bibfnamefont{K.}~\bibnamefont{Segawa}}, \bibnamefont{and}
  \bibinfo{author}{\bibfnamefont{Y.}~\bibnamefont{Ando}}, \bibinfo{journal}{Nat
  Phys} \textbf{\bibinfo{volume}{8}}, \bibinfo{pages}{800}
  (\bibinfo{year}{2012}{\natexlab{b}}).

\bibitem[{\citenamefont{Dziawa et~al.}(2012)\citenamefont{Dziawa, Kowalski,
  Dybko, Buczko, Szczerbakow, Szot, {\L}usakowska, Balasubramanian, Wojek,
  Berntsen et~al.}}]{PbSnSe_ARPES}
\bibinfo{author}{\bibfnamefont{P.}~\bibnamefont{Dziawa}},
  \bibinfo{author}{\bibfnamefont{B.~J.} \bibnamefont{Kowalski}},
  \bibinfo{author}{\bibfnamefont{K.}~\bibnamefont{Dybko}},
  \bibinfo{author}{\bibfnamefont{R.}~\bibnamefont{Buczko}},
  \bibinfo{author}{\bibfnamefont{A.}~\bibnamefont{Szczerbakow}},
  \bibinfo{author}{\bibfnamefont{M.}~\bibnamefont{Szot}},
  \bibinfo{author}{\bibfnamefont{E.}~\bibnamefont{{\L}usakowska}},
  \bibinfo{author}{\bibfnamefont{T.}~\bibnamefont{Balasubramanian}},
  \bibinfo{author}{\bibfnamefont{B.~M.} \bibnamefont{Wojek}},
  \bibinfo{author}{\bibfnamefont{M.~H.} \bibnamefont{Berntsen}},
  \bibnamefont{et~al.}, \bibinfo{journal}{Nat Mater}
  \textbf{\bibinfo{volume}{11}}, \bibinfo{pages}{1023} (\bibinfo{year}{2012}).

\bibitem[{\citenamefont{Xu et~al.}(2012)\citenamefont{Xu, Liu, Alidoust,
  Neupane, Qian, Belopolski, Denlinger, Wang, Lin, Wray et~al.}}]{PbSnTe_ARPES}
\bibinfo{author}{\bibfnamefont{S.-Y.} \bibnamefont{Xu}},
  \bibinfo{author}{\bibfnamefont{C.}~\bibnamefont{Liu}},
  \bibinfo{author}{\bibfnamefont{N.}~\bibnamefont{Alidoust}},
  \bibinfo{author}{\bibfnamefont{M.}~\bibnamefont{Neupane}},
  \bibinfo{author}{\bibfnamefont{D.}~\bibnamefont{Qian}},
  \bibinfo{author}{\bibfnamefont{I.}~\bibnamefont{Belopolski}},
  \bibinfo{author}{\bibfnamefont{J.}~\bibnamefont{Denlinger}},
  \bibinfo{author}{\bibfnamefont{Y.}~\bibnamefont{Wang}},
  \bibinfo{author}{\bibfnamefont{H.}~\bibnamefont{Lin}},
  \bibinfo{author}{\bibfnamefont{L.}~\bibnamefont{Wray}}, \bibnamefont{et~al.},
  \bibinfo{journal}{Nat Commun} \textbf{\bibinfo{volume}{3}},
  \bibinfo{pages}{1192} (\bibinfo{year}{2012}).

\bibitem[{\citenamefont{Liu et~al.}(2014)\citenamefont{Liu, Hsieh, Wei, Duan,
  Moodera, and Fu}}]{TCI_app}
\bibinfo{author}{\bibfnamefont{J.}~\bibnamefont{Liu}},
  \bibinfo{author}{\bibfnamefont{T.~H.} \bibnamefont{Hsieh}},
  \bibinfo{author}{\bibfnamefont{P.}~\bibnamefont{Wei}},
  \bibinfo{author}{\bibfnamefont{W.}~\bibnamefont{Duan}},
  \bibinfo{author}{\bibfnamefont{J.}~\bibnamefont{Moodera}}, \bibnamefont{and}
  \bibinfo{author}{\bibfnamefont{L.}~\bibnamefont{Fu}}, \bibinfo{journal}{Nat
  Mater} \textbf{\bibinfo{volume}{13}}, \bibinfo{pages}{178}
  (\bibinfo{year}{2014}).

\bibitem[{\citenamefont{Fang et~al.}(2014)\citenamefont{Fang, Gilbert, and
  Bernevig}}]{PhysRevLett.112.046801}
\bibinfo{author}{\bibfnamefont{C.}~\bibnamefont{Fang}},
  \bibinfo{author}{\bibfnamefont{M.~J.} \bibnamefont{Gilbert}},
  \bibnamefont{and} \bibinfo{author}{\bibfnamefont{B.~A.}
  \bibnamefont{Bernevig}}, \bibinfo{journal}{Phys. Rev. Lett.}
  \textbf{\bibinfo{volume}{112}}, \bibinfo{pages}{046801}
  (\bibinfo{year}{2014}).

\bibitem[{\citenamefont{Zeljkovic et~al.}(2015)\citenamefont{Zeljkovic, Okada,
  Serbyn, Sankar, Walkup, Zhou, Liu, Chang, Wang, Hasan
  et~al.}}]{TCI_app_strain}
\bibinfo{author}{\bibfnamefont{I.}~\bibnamefont{Zeljkovic}},
  \bibinfo{author}{\bibfnamefont{Y.}~\bibnamefont{Okada}},
  \bibinfo{author}{\bibfnamefont{M.}~\bibnamefont{Serbyn}},
  \bibinfo{author}{\bibfnamefont{R.}~\bibnamefont{Sankar}},
  \bibinfo{author}{\bibfnamefont{D.}~\bibnamefont{Walkup}},
  \bibinfo{author}{\bibfnamefont{W.}~\bibnamefont{Zhou}},
  \bibinfo{author}{\bibfnamefont{J.}~\bibnamefont{Liu}},
  \bibinfo{author}{\bibfnamefont{G.}~\bibnamefont{Chang}},
  \bibinfo{author}{\bibfnamefont{Y.~J.} \bibnamefont{Wang}},
  \bibinfo{author}{\bibfnamefont{M.~Z.} \bibnamefont{Hasan}},
  \bibnamefont{et~al.}, \bibinfo{journal}{Nat Mater}
  \textbf{\bibinfo{volume}{14}}, \bibinfo{pages}{318} (\bibinfo{year}{2015}).

\bibitem[{\citenamefont{Bushmarina et~al.}(1986)\citenamefont{Bushmarina,
  Drabkin, Kompaniets, Parfenfev, and Shakhov}}]{InSnTe_SC}
\bibinfo{author}{\bibfnamefont{G.~S.} \bibnamefont{Bushmarina}},
  \bibinfo{author}{\bibfnamefont{I.~A.} \bibnamefont{Drabkin}},
  \bibinfo{author}{\bibfnamefont{V.}~\bibnamefont{Kompaniets}},
  \bibinfo{author}{\bibfnamefont{R.}~\bibnamefont{Parfenfev}},
  \bibnamefont{and} \bibinfo{author}{\bibfnamefont{M.~A.}
  \bibnamefont{Shakhov}}, \bibinfo{journal}{Sov. Phys. Solid State}
  \textbf{\bibinfo{volume}{28}}, \bibinfo{pages}{612} (\bibinfo{year}{1986}).

\bibitem[{\citenamefont{Sasaki et~al.}(2012)\citenamefont{Sasaki, Ren, Taskin,
  Segawa, Fu, and Ando}}]{PhysRevLett.109.217004}
\bibinfo{author}{\bibfnamefont{S.}~\bibnamefont{Sasaki}},
  \bibinfo{author}{\bibfnamefont{Z.}~\bibnamefont{Ren}},
  \bibinfo{author}{\bibfnamefont{A.~A.} \bibnamefont{Taskin}},
  \bibinfo{author}{\bibfnamefont{K.}~\bibnamefont{Segawa}},
  \bibinfo{author}{\bibfnamefont{L.}~\bibnamefont{Fu}}, \bibnamefont{and}
  \bibinfo{author}{\bibfnamefont{Y.}~\bibnamefont{Ando}},
  \bibinfo{journal}{Phys. Rev. Lett.} \textbf{\bibinfo{volume}{109}},
  \bibinfo{pages}{217004} (\bibinfo{year}{2012}).

\bibitem[{\citenamefont{Sato et~al.}(2013)\citenamefont{Sato, Tanaka, Nakayama,
  Souma, Takahashi, Sasaki, Ren, Taskin, Segawa, and
  Ando}}]{PhysRevLett.110.206804}
\bibinfo{author}{\bibfnamefont{T.}~\bibnamefont{Sato}},
  \bibinfo{author}{\bibfnamefont{Y.}~\bibnamefont{Tanaka}},
  \bibinfo{author}{\bibfnamefont{K.}~\bibnamefont{Nakayama}},
  \bibinfo{author}{\bibfnamefont{S.}~\bibnamefont{Souma}},
  \bibinfo{author}{\bibfnamefont{T.}~\bibnamefont{Takahashi}},
  \bibinfo{author}{\bibfnamefont{S.}~\bibnamefont{Sasaki}},
  \bibinfo{author}{\bibfnamefont{Z.}~\bibnamefont{Ren}},
  \bibinfo{author}{\bibfnamefont{A.~A.} \bibnamefont{Taskin}},
  \bibinfo{author}{\bibfnamefont{K.}~\bibnamefont{Segawa}}, \bibnamefont{and}
  \bibinfo{author}{\bibfnamefont{Y.}~\bibnamefont{Ando}},
  \bibinfo{journal}{Phys. Rev. Lett.} \textbf{\bibinfo{volume}{110}},
  \bibinfo{pages}{206804} (\bibinfo{year}{2013}).

\bibitem[{\citenamefont{Lent et~al.}(1986)\citenamefont{Lent, Bowen, Dow,
  Alligaier, Sankey, and Ho}}]{model}
\bibinfo{author}{\bibfnamefont{A.~C.} \bibnamefont{Lent}},
  \bibinfo{author}{\bibfnamefont{M.~S.} \bibnamefont{Bowen}},
  \bibinfo{author}{\bibfnamefont{J.~D.} \bibnamefont{Dow}},
  \bibinfo{author}{\bibfnamefont{R.~S.} \bibnamefont{Alligaier}},
  \bibinfo{author}{\bibfnamefont{O.~F.} \bibnamefont{Sankey}},
  \bibnamefont{and} \bibinfo{author}{\bibfnamefont{E.~S.} \bibnamefont{Ho}},
  \bibinfo{journal}{Superlattices Microstruct} \textbf{\bibinfo{volume}{2}},
  \bibinfo{pages}{491} (\bibinfo{year}{1986}).

\bibitem[{\citenamefont{Novak et~al.}(2013)\citenamefont{Novak, Sasaki,
  Kriener, Segawa, and Ando}}]{PhysRevB.88.140502}
\bibinfo{author}{\bibfnamefont{M.}~\bibnamefont{Novak}},
  \bibinfo{author}{\bibfnamefont{S.}~\bibnamefont{Sasaki}},
  \bibinfo{author}{\bibfnamefont{M.}~\bibnamefont{Kriener}},
  \bibinfo{author}{\bibfnamefont{K.}~\bibnamefont{Segawa}}, \bibnamefont{and}
  \bibinfo{author}{\bibfnamefont{Y.}~\bibnamefont{Ando}},
  \bibinfo{journal}{Phys. Rev. B} \textbf{\bibinfo{volume}{88}},
  \bibinfo{pages}{140502} (\bibinfo{year}{2013}).

\bibitem[{\citenamefont{Erickson et~al.}(2009)\citenamefont{Erickson, Chu,
  Toney, Geballe, and Fisher}}]{PhysRevB.79.024520}
\bibinfo{author}{\bibfnamefont{A.~S.} \bibnamefont{Erickson}},
  \bibinfo{author}{\bibfnamefont{J.-H.} \bibnamefont{Chu}},
  \bibinfo{author}{\bibfnamefont{M.~F.} \bibnamefont{Toney}},
  \bibinfo{author}{\bibfnamefont{T.~H.} \bibnamefont{Geballe}},
  \bibnamefont{and} \bibinfo{author}{\bibfnamefont{I.~R.}
  \bibnamefont{Fisher}}, \bibinfo{journal}{Phys. Rev. B}
  \textbf{\bibinfo{volume}{79}}, \bibinfo{pages}{024520}
  (\bibinfo{year}{2009}).

\bibitem[{\citenamefont{Balakrishnan et~al.}(2013)\citenamefont{Balakrishnan,
  Bawden, Cavendish, and Lees}}]{PhysRevB.87.140507}
\bibinfo{author}{\bibfnamefont{G.}~\bibnamefont{Balakrishnan}},
  \bibinfo{author}{\bibfnamefont{L.}~\bibnamefont{Bawden}},
  \bibinfo{author}{\bibfnamefont{S.}~\bibnamefont{Cavendish}},
  \bibnamefont{and} \bibinfo{author}{\bibfnamefont{M.~R.} \bibnamefont{Lees}},
  \bibinfo{journal}{Phys. Rev. B} \textbf{\bibinfo{volume}{87}},
  \bibinfo{pages}{140507} (\bibinfo{year}{2013}).

\bibitem[{\citenamefont{Saghir et~al.}(2014)\citenamefont{Saghir, Barker,
  Balakrishnan, Hillier, and Lees}}]{PhysRevB.90.064508}
\bibinfo{author}{\bibfnamefont{M.}~\bibnamefont{Saghir}},
  \bibinfo{author}{\bibfnamefont{J.~A.~T.} \bibnamefont{Barker}},
  \bibinfo{author}{\bibfnamefont{G.}~\bibnamefont{Balakrishnan}},
  \bibinfo{author}{\bibfnamefont{A.~D.} \bibnamefont{Hillier}},
  \bibnamefont{and} \bibinfo{author}{\bibfnamefont{M.~R.} \bibnamefont{Lees}},
  \bibinfo{journal}{Phys. Rev. B} \textbf{\bibinfo{volume}{90}},
  \bibinfo{pages}{064508} (\bibinfo{year}{2014}).

\bibitem[{\citenamefont{He et~al.}(2013)\citenamefont{He, Zhang, Pan, Hong,
  Zhou, and Li}}]{PhysRevB.88.014523}
\bibinfo{author}{\bibfnamefont{L.~P.} \bibnamefont{He}},
  \bibinfo{author}{\bibfnamefont{Z.}~\bibnamefont{Zhang}},
  \bibinfo{author}{\bibfnamefont{J.}~\bibnamefont{Pan}},
  \bibinfo{author}{\bibfnamefont{X.~C.} \bibnamefont{Hong}},
  \bibinfo{author}{\bibfnamefont{S.~Y.} \bibnamefont{Zhou}}, \bibnamefont{and}
  \bibinfo{author}{\bibfnamefont{S.~Y.} \bibnamefont{Li}},
  \bibinfo{journal}{Phys. Rev. B} \textbf{\bibinfo{volume}{88}},
  \bibinfo{pages}{014523} (\bibinfo{year}{2013}).

\bibitem[{\citenamefont{Sigrist and Ueda}(1991)}]{RevModPhys.63.239}
\bibinfo{author}{\bibfnamefont{M.}~\bibnamefont{Sigrist}} \bibnamefont{and}
  \bibinfo{author}{\bibfnamefont{K.}~\bibnamefont{Ueda}},
  \bibinfo{journal}{Rev. Mod. Phys.} \textbf{\bibinfo{volume}{63}},
  \bibinfo{pages}{239} (\bibinfo{year}{1991}).

\bibitem[{\citenamefont{Ueda and Rice}(1985)}]{PhysRevB.31.7114}
\bibinfo{author}{\bibfnamefont{K.}~\bibnamefont{Ueda}} \bibnamefont{and}
  \bibinfo{author}{\bibfnamefont{T.~M.} \bibnamefont{Rice}},
  \bibinfo{journal}{Phys. Rev. B} \textbf{\bibinfo{volume}{31}},
  \bibinfo{pages}{7114} (\bibinfo{year}{1985}).

\bibitem[{\citenamefont{Blount}(1985)}]{PhysRevB.32.2935}
\bibinfo{author}{\bibfnamefont{E.~I.} \bibnamefont{Blount}},
  \bibinfo{journal}{Phys. Rev. B} \textbf{\bibinfo{volume}{32}},
  \bibinfo{pages}{2935} (\bibinfo{year}{1985}).

\bibitem[{\citenamefont{Anderson}(1984)}]{PhysRevB.30.4000}
\bibinfo{author}{\bibfnamefont{P.~W.} \bibnamefont{Anderson}},
  \bibinfo{journal}{Phys. Rev. B} \textbf{\bibinfo{volume}{30}},
  \bibinfo{pages}{4000} (\bibinfo{year}{1984}).

\bibitem[{\citenamefont{Chiu et~al.}(2013)\citenamefont{Chiu, Yao, and
  Ryu}}]{PhysRevB.88.075142}
\bibinfo{author}{\bibfnamefont{C.-K.} \bibnamefont{Chiu}},
  \bibinfo{author}{\bibfnamefont{H.}~\bibnamefont{Yao}}, \bibnamefont{and}
  \bibinfo{author}{\bibfnamefont{S.}~\bibnamefont{Ryu}},
  \bibinfo{journal}{Phys. Rev. B} \textbf{\bibinfo{volume}{88}},
  \bibinfo{pages}{075142} (\bibinfo{year}{2013}).

\bibitem[{\citenamefont{Shiozaki and Sato}(2014)}]{PhysRevB.90.165114}
\bibinfo{author}{\bibfnamefont{K.}~\bibnamefont{Shiozaki}} \bibnamefont{and}
  \bibinfo{author}{\bibfnamefont{M.}~\bibnamefont{Sato}},
  \bibinfo{journal}{Phys. Rev. B} \textbf{\bibinfo{volume}{90}},
  \bibinfo{pages}{165114} (\bibinfo{year}{2014}).

\bibitem[{\citenamefont{Morimoto and Furusaki}(2013)}]{PhysRevB.88.125129}
\bibinfo{author}{\bibfnamefont{T.}~\bibnamefont{Morimoto}} \bibnamefont{and}
  \bibinfo{author}{\bibfnamefont{A.}~\bibnamefont{Furusaki}},
  \bibinfo{journal}{Phys. Rev. B} \textbf{\bibinfo{volume}{88}},
  \bibinfo{pages}{125129} (\bibinfo{year}{2013}).

\bibitem[{\citenamefont{Altland and Zirnbauer}(1997)}]{PhysRevB.55.1142}
\bibinfo{author}{\bibfnamefont{A.}~\bibnamefont{Altland}} \bibnamefont{and}
  \bibinfo{author}{\bibfnamefont{M.~R.} \bibnamefont{Zirnbauer}},
  \bibinfo{journal}{Phys. Rev. B} \textbf{\bibinfo{volume}{55}},
  \bibinfo{pages}{1142} (\bibinfo{year}{1997}).

\bibitem[{\citenamefont{Kitaev}(2009)}]{Kitaev2}
\bibinfo{author}{\bibfnamefont{A.}~\bibnamefont{Kitaev}}, \bibinfo{journal}{AIP
  Conference Proceedings} \textbf{\bibinfo{volume}{1134}}, \bibinfo{pages}{22}
  (\bibinfo{year}{2009}).

\bibitem[{\citenamefont{Ryu et~al.}(2010)\citenamefont{Ryu, Schnyder, Furusaki,
  and Ludwig}}]{Schnyder}
\bibinfo{author}{\bibfnamefont{S.}~\bibnamefont{Ryu}},
  \bibinfo{author}{\bibfnamefont{A.~P.} \bibnamefont{Schnyder}},
  \bibinfo{author}{\bibfnamefont{A.}~\bibnamefont{Furusaki}}, \bibnamefont{and}
  \bibinfo{author}{\bibfnamefont{A.~W.~W.} \bibnamefont{Ludwig}},
  \bibinfo{journal}{New Journal of Physics} \textbf{\bibinfo{volume}{12}},
  \bibinfo{pages}{065010} (\bibinfo{year}{2010}).

\bibitem[{\citenamefont{Schnyder et~al.}(2008)\citenamefont{Schnyder, Ryu,
  Furusaki, and Ludwig}}]{PhysRevB.78.195125}
\bibinfo{author}{\bibfnamefont{A.~P.} \bibnamefont{Schnyder}},
  \bibinfo{author}{\bibfnamefont{S.}~\bibnamefont{Ryu}},
  \bibinfo{author}{\bibfnamefont{A.}~\bibnamefont{Furusaki}}, \bibnamefont{and}
  \bibinfo{author}{\bibfnamefont{A.~W.~W.} \bibnamefont{Ludwig}},
  \bibinfo{journal}{Phys. Rev. B} \textbf{\bibinfo{volume}{78}},
  \bibinfo{pages}{195125} (\bibinfo{year}{2008}).

\bibitem[{\citenamefont{Sato}(2010)}]{PhysRevB.81.220504}
\bibinfo{author}{\bibfnamefont{M.}~\bibnamefont{Sato}}, \bibinfo{journal}{Phys.
  Rev. B} \textbf{\bibinfo{volume}{81}}, \bibinfo{pages}{220504}
  (\bibinfo{year}{2010}).

\bibitem[{\citenamefont{Fukui et~al.}(2005)\citenamefont{Fukui, Hatsugai, and
  Suzuki}}]{doi:10.1143/JPSJ.74.1674}
\bibinfo{author}{\bibfnamefont{T.}~\bibnamefont{Fukui}},
  \bibinfo{author}{\bibfnamefont{Y.}~\bibnamefont{Hatsugai}}, \bibnamefont{and}
  \bibinfo{author}{\bibfnamefont{H.}~\bibnamefont{Suzuki}},
  \bibinfo{journal}{Journal of the Physical Society of Japan}
  \textbf{\bibinfo{volume}{74}}, \bibinfo{pages}{1674} (\bibinfo{year}{2005}).

\bibitem[{\citenamefont{Ueno et~al.}(2013)\citenamefont{Ueno, Yamakage, Tanaka,
  and Sato}}]{PhysRevLett.111.087002}
\bibinfo{author}{\bibfnamefont{Y.}~\bibnamefont{Ueno}},
  \bibinfo{author}{\bibfnamefont{A.}~\bibnamefont{Yamakage}},
  \bibinfo{author}{\bibfnamefont{Y.}~\bibnamefont{Tanaka}}, \bibnamefont{and}
  \bibinfo{author}{\bibfnamefont{M.}~\bibnamefont{Sato}},
  \bibinfo{journal}{Phys. Rev. Lett.} \textbf{\bibinfo{volume}{111}},
  \bibinfo{pages}{087002} (\bibinfo{year}{2013}).

\bibitem[{\citenamefont{Umerski}(1997)}]{PhysRevB.55.5266}
\bibinfo{author}{\bibfnamefont{A.}~\bibnamefont{Umerski}},
  \bibinfo{journal}{Phys. Rev. B} \textbf{\bibinfo{volume}{55}},
  \bibinfo{pages}{5266} (\bibinfo{year}{1997}).

\bibitem[{\citenamefont{Ozawa et~al.}(2014)\citenamefont{Ozawa, Yamakage, Sato,
  and Tanaka}}]{PhysRevB.90.045309}
\bibinfo{author}{\bibfnamefont{H.}~\bibnamefont{Ozawa}},
  \bibinfo{author}{\bibfnamefont{A.}~\bibnamefont{Yamakage}},
  \bibinfo{author}{\bibfnamefont{M.}~\bibnamefont{Sato}}, \bibnamefont{and}
  \bibinfo{author}{\bibfnamefont{Y.}~\bibnamefont{Tanaka}},
  \bibinfo{journal}{Phys. Rev. B} \textbf{\bibinfo{volume}{90}},
  \bibinfo{pages}{045309} (\bibinfo{year}{2014}).

\bibitem[{\citenamefont{Goswami and Roy}(2014)}]{PhysRevB.90.041301}
\bibinfo{author}{\bibfnamefont{P.}~\bibnamefont{Goswami}} \bibnamefont{and}
  \bibinfo{author}{\bibfnamefont{B.}~\bibnamefont{Roy}},
  \bibinfo{journal}{Phys. Rev. B} \textbf{\bibinfo{volume}{90}},
  \bibinfo{pages}{041301} (\bibinfo{year}{2014}).

\end{thebibliography}

\end{document}